\documentclass[11pt]{article}
\usepackage{jheppub}
\usepackage{braket} 

\usepackage{tikz}
\usetikzlibrary{intersections, calc,positioning,decorations.pathreplacing,decorations.pathmorphing,arrows,arrows.meta,patterns}

\usepackage{amsthm}

\newtheorem*{conjecture}{Conjecture}

\def\tr{{\rm tr}}
\def\Tr{{\rm Tr}}
\def\d{{\rm d}}
\def\i{{\rm i}}

\def\CB{{\cal B}}

\def\CD{{\cal D}}

\def\CK{{\cal K}}
\def\CL{{\cal L}}
\def\CR{{\cal R}}
\def\CM{{\cal M}}
\def\CN{{\cal N}}
\def\CO{{\cal O}}
\def\CP{{\cal P}}
\def\CQ{{\cal Q}}

\def\CT{{\cal T}}

\def\BC{\mathbb{C}}

\def\BH{\mathbb{H}}
\def\BR{\mathbb{R}}
\def\BS{\mathbb{S}}

\def\d{\mathrm{d}}
\def\SO{\mathrm{SO}}
\def\SU{\mathrm{SU}}

\def\U{\mathrm{U}}

\newcommand{\dd}{\mathrm{d}}

\title{Towards a $C$-theorem in defect CFT}

\author[a,b]{Nozomu Kobayashi,}
\author[a]{Tatsuma Nishioka,}
\author[a]{Yoshiki Sato}
\author[a]{and Kento Watanabe}

\affiliation[a]{Department of Physics, Faculty of Science,
The University of Tokyo,\\
Bunkyo-ku, Tokyo 113-0033, Japan}
\affiliation[b]{Kavli Institute for the Physics and Mathematics of the Universe (WPI),\\ 
The University of Tokyo Institutes for Advanced Study, The University of Tokyo,\\
Kashiwa, Chiba 277-8583, Japan}

\abstract{
We explore a $C$-theorem in defect conformal field theories (DCFTs) that unify all the known conjectures and theorems until now.
We examine as a candidate $C$-function the additional contributions from conformal defects to the sphere free energy and the entanglement entropy across a sphere in a number of examples including holographic models.
We find the two quantities are equivalent, when suitably regularized, for codimension-one defects (or boundaries), but differ by a universal constant term otherwise.
Moreover, we find in a few field theoretic examples that the sphere free energy decreases but the entanglement entropy increases along a certain renormalization group (RG) flow triggered by a defect localized perturbation which is assumed to have a trivial IR fixed point without defects.
We hence propose a $C$-theorem in DCFTs stating that the increment of the regularized sphere free energy due to the defect does not increase under any defect RG flow.
We also provide a proof of our proposal in several holographic models of defect RG flows.
}

\preprint{UT-18-22, IPMU18-0166}

\begin{document}
\maketitle

\section{Introduction}

Understanding the dynamics under a renormalization group (RG) flow is central to the studies of quantum field theories (QFTs).
Among the most challenging problems is proving the irreversibility of the RG flow, which is quantitatively guaranteed by the existence of a monotonically decreasing function $C (\lambda)$ interpolating between two theories parametrized by a set of coupling constants $\lambda = (\lambda_1, \lambda_2,\cdots)$ in the theory space.
A $C$-function is regarded as a measure counting the effective degrees of freedom in QFT, and the monotonicity, called the $C$-theorem, provides non-perturbative constraints on the RG dynamics that are inaccessible by other means.

A typical class of RG flows is a deformation of a conformal field theory (CFT)
\begin{align}
I_{\text{CFT}}+\lambda  \int \! \dd^d x\, \sqrt{g} \,  \mathcal{O}(x) \ ,
\end{align}
by a relevant operator $\mathcal{O}$ of dimension $\Delta \le d$, which triggers a flow from the UV CFT to another CFT at the IR fixed point.
In $d=2$, there exists a $C$-function that decreases under the RG flow monotonically and coincides with the central charges at the conformal fixed points \cite{Zamolodchikov:1986gt}.
In higher even dimensions, the type $A$ central charge of the conformal anomaly is speculated to be a $C$-function \cite{Cardy:1988cwa,Myers:2010xs,Myers:2010tj}, and a proof has been established in $d=4$ \cite{Komargodski:2011vj}.
On the other hand, there are no conformal anomalies in odd dimensions, but
it was conjectured that the sphere free energy $F \equiv (-1)^{(d-1)/2}\,\log\,Z[\BS^d]$, defined by the conformal invariant partition function $Z[\BS^d]$ of CFT on a $d$-sphere, be monotonic under any RG flow \cite{Jafferis:2011zi,Klebanov:2011gs}.
The conjecture has been extended to continuous $d$ dimensions as the generalized $F$-theorem \cite{Giombi:2014xxa} by interpolating between the type $A$ anomaly in even $d$ and the sphere free energy $F$ in odd $d$, resulting in the statement that (the universal part of) the quantity
\begin{align}\label{Generalized_F}
	\tilde F \equiv \sin \left( \frac{\pi\,d}{2} \right)\,\log\,Z[\BS^d] \ ,
\end{align}
is positive and does not increase along any RG flow
\begin{align}\label{Generalized_F_theorem}
	\tilde F_\text{UV} \ge \tilde F_\text{IR} \ .
\end{align}
This is one of the most general $C$-theorems proposed in arbitrary dimensions so far.\footnote{See \cite{Kawano:2014moa} for a holographic proof of the generalized $F$-theorem.}

Another approach to establishing a $C$-theorem is to use the entanglement entropy across an entangling surface $\Sigma$ dividing the spacial slice into two regions at a constant time.
In $d$-dimensional QFTs, the entanglement entropy takes the following general form,
\begin{align}
S^\text{(CFT)} =\frac{A_{d-2}}{\epsilon^{d-2}}+\frac{A_{d-4}}{\epsilon^{d-4}}+\cdots +
\begin{cases}
a_{\log} \log \left( \frac{R}{\epsilon} \right)\ , & (d=\text{even}) \ , \\
a_0\ , &(d=\text{odd})\ ,
\end{cases}
\label{ee}
\end{align}
where $R$ is a typical size of the entangling region and $\epsilon\ll R$ is the UV cutoff .
When the theory is conformal and $\Sigma$ is spherical the constants $a_{\log}, a_0$ are universal in the sense that they are independent of the regularization scheme (i.e., the choice of the UV cutoff $\epsilon$), and are conjectured to be $C$-functions in even and odd dimensions respectively \cite{Myers:2010xs,Myers:2010tj}.
This entropic version of the $C$-theorem looks quite different from the generalized $F$-theorem based on the sphere free energy \eqref{Generalized_F}, but the two statements turn out to be the same due to the striking relation
\begin{align}\label{CHM_relation}
	S^\text{(CFT)} = \log\, Z^\text{(CFT)} \ ,
\end{align}
which holds, as the superscripts indicate, for CFT with a spherical entangling surface up to UV divergences  \cite{Casini:2011kv}.
This equivalence is actually the key to proving the $F$-theorem in $d=3$ \cite{Casini:2012ei} where the monotonic property of entanglement entropy was adapted to show the monotonicity of a function built from the entanglement entropy across a sphere \cite{Liu:2012eea} which interpolates the sphere free energies at the UV and IR fixed points.
Extending the proof in $d=3$ (and $d=2$ \cite{Casini:2004bw}) to higher dimensions was attempted in \cite{Casini:2017vbe}, which amounts to a different monotonicity theorem from \eqref{Generalized_F_theorem} in $d>4$ (see also \cite{Lashkari:2017rcl}).
It still remains open whether the $F$-theorem \eqref{Generalized_F_theorem} in higher dimensions follows from the quantum inequality of a certain entanglement measure.

Instead of going to higher dimensions one can introduce a boundary to the spacetime or extended objects called \emph{defects} to QFTs, and ask if a certain type of $C$-theorems remain to hold even in such cases.
In the former case, a partial answer to this question is known as the $g$-theorem in two-dimensional boundary CFTs (BCFTs) \cite{Affleck:1991tk,Friedan:2003yc}, which states that 
the $g$-function, the constant term of the thermal entropy independent of the system size, monotonically decreases under a boundary RG flow, an RG flow triggered by the relevant perturbation localized on the boundary. 
The $g$-theorem also has an alternative proof that relies on the equivalence of the $g$-function and the boundary entropy,
the difference of the entanglement entropies between BCFT and CFT,
\begin{align}
S_{\text{bdy}}=S^{\text{(BCFT)}}-\frac{1}{2}\,S^{\text{(CFT)}} \ ,
\end{align}
at the conformal fixed point.
The monotonicity of the $g$-function is shown to follow from the positivity of the relative entropy \cite{Casini:2016fgb}. 
In higher-dimensional BCFTs, there are several proposals for $g$-functions, the hemisphere partition function (the boundary $F$-theorem) \cite{Nozaki:2012qd,Gaiotto:2014gha}, the boundary entropy \cite{Estes:2014hka} and the holographic $g$-functions \cite{Yamaguchi:2002pa,Takayanagi:2011zk,Fujita:2011fp} with varying degrees of evidences.\footnote{The holographic $g$-functions are proven to be monotonic under any holographic boundary RG flow satisfying the null energy condition, but their physical meanings are unclear unless the theory is at the fixed point as in the case of the holographic $c$-theorem \cite{Girardello:1998pd,Freedman:1999gp}.}
The first two proposals are not independent but the same statement as an analogous identity to \eqref{CHM_relation} holds for BCFT.\footnote{The partition function of BCFT is defined on a hemisphere $\BH\BS^d$, so $Z^\text{(BCFT)}\equiv Z [\BH\BS^d]$.}

Moving onto the case with defects we focus on $d$-dimensional defect CFT (DCFT) with a planar or spherical conformal defect of dimension $p$ that we denote by $\CD^{(p)}$ preserving the maximal subgroup of the conformal group.
In this case, less is known for $C$-theorems except a rigorous result for two-dimensional defects or boundaries \cite{JO1}.
This is called the $b$-theorem stating the monotonicity of the coefficient $b$ of the Weyl anomaly on the submanifolds of dimension $p=2$
\begin{align}
b_{\text{UV}}\geq b_{\text{IR}}\ .
\end{align}
When the ambient spacetime is three-dimensional ($d=3$), the $b$-theorem implies the $g$-theorem in BCFT$_3$.
For $d>3$ it yields a class of $C$-theorems in DCFTs.
See table \ref{table1} for the summary of the current status.

In this paper we explore a $C$-theorem in DCFTs with defects of various dimensions.
Namely we want to establish a monotonically decreasing function under a defect RG flow triggered by relevantly perturbing a DCFT
\begin{align}
I=I_{\mathrm{DCFT}}+\hat{\lambda} \int \! \dd^p \hat x\, \sqrt{\hat{g}} \,  \hat{\mathcal{O}}(\hat{x}) \ .
\end{align}
An important guiding principle for the search is that the candidate $C$-theorem should reproduce all the known conjectures and theorems in the appropriate limits.
We are then left with two possibilities, the defect free energy, the additional contribution to the sphere free energy from the spherical defect
\begin{align}
	\log\,\langle\, \CD^{(p)}\,\rangle = \log\, Z^\text{(DCFT)} - \log\, Z^\text{(CFT)} \ ,
\end{align}
or the defect entropy, the increment of the entanglement entropy across a sphere due to the planer defect\footnote{The defect entropy has been conjectured to be a $C$-function for interface CFTs in \cite{Estes:2014hka} based on the studies of several holographic models.
}
\begin{align}
S_{\text{defect}}=S^{\text{(DCFT)}}-S^{\text{(CFT)}} \ .
\end{align}
These are expected to count the degrees of freedom on the defect, but have UV divergent terms that need to be regularized and renormalized so as to be a well-defined $C$-function.
After the regularization we find the resulting quantities are universal, i.e., do not depend on the regularization scheme when evaluated at the conformal fixed point.

Reminding the relation \eqref{CHM_relation} one may suspect a similar identity holds between the defect free energy and the defect entropy.
Indeed they are equivalent up to UV divergences when $p=d-1$, but differ by a term fixed by the one-point function of the stress-energy tensor for $p< d-1$.
Their precise relation is derived in \eqref{Defect_Entropy_universal} by using the conformal transformation known as the Casini-Huerta-Myers (CHM) map \cite{Casini:2011kv,Jensen:2013lxa}.

We study a variety of examples of DCFTs and the holographic models with the hope of finding defect RG flows that exclude the possibility of one of the two being a $C$-function.
Most of our field theoretic examples are DCFTs with line defects and we assume that the theories are connected to the trivial fixed points, i.e., the ambient CFTs without defects by certain defect RG flows.
We find several examples where the defect free energy decreases but the defect entropy increases along the flow.
On the other hand both of them always decrease in all the holographic models we study.
These observations therefore lead us to propose a $C$-theorem in DCFTs stating that the universal part of the defect free energy 
\begin{align}
	\tilde D \equiv \sin\left(\frac{\pi\,p}{2}\right)\,\log\,\langle\, \CD^{(p)}\,\rangle \ ,
\end{align}
decreases along any defect RG flow
\begin{align}
	\tilde D_\text{UV} \ge \tilde D_\text{IR} \ .
\end{align}
The more precise statement is presented around \eqref{D_inequality}.
Note that this should be seen as the counterpart to the generalized $F$-theorem \eqref{Generalized_F_theorem} in CFTs.
In fact it reduces to the generalized $F$-theorem on the defect when the defect theory decouples from the ambient theory.
Moreover our proposal unifies the higher-dimensional $g$-theorems and the $b$-theorem for $p=d-1$ and $p=2$ respectively, and asserts a new family of $C$-theorems otherwise (see table \ref{table1} for the summary).

\begin{table}[tt]
\centering
\begin{tikzpicture}
\filldraw[fill=cyan!40,line width=1pt]
(12.8,0)--(0,0)--(0,-2)--(3.2,-2)--(3.2,-4)--(6.4,-4)--(6.4,-6)--(9.6,-6)--(9.6,-8)--(12.8,-8){};
\filldraw[fill=gray!20, draw=white, line width=1pt, rounded corners=10pt]
(0.1,-0.1)--(3.1,-0.1)--(3.1,-1.9)--(0.1,-1.9)-- cycle;
\filldraw[fill=gray!20, draw=white, line width=1pt,  rounded corners=10pt]
(12.8,-2.1)--(3.3,-2.1)--(3.3,-3.9)--(12.8,-3.9);
\filldraw[fill=gray!20, draw=white, line width=1pt,  rounded corners=10pt]
(6.5,-4.1)--(9.5,-4.1)--(9.5,-5.9)--(6.5,-5.9)-- cycle;

\node[]  at (1.6,0.5) {$d=2$};
\node[]  at (4.8,0.5) {$d=3$};
\node[]  at (8,0.5) {$d=4$};
\node[]  at (11.2,0.5) {$d=5$};
\node[]  at (-0.8,-1) {$p=1$};
\node[]  at (2.4,-3) {$p=2$};
\node[]  at (5.6,-5) {$p=3$};
\node[]  at (8.8,-7) {$p=4$};

\node[text width=3cm,align=center]  at (1.6,-1) {$g$-theorem\\Proof \cite{Friedan:2003yc,Casini:2016fgb}};
\node[text width=3cm,align=center]  at (8,-3) {$b$-theorem\\Proof \cite{JO1}};
\node[text width=3cm,align=center]  at (8,-5) {bdy $F$-theorem\\Proposal \cite{Nozaki:2012qd,Gaiotto:2014gha}};
\end{tikzpicture}
\caption{Summary of the conjectured and proved $C$-theorems in BCFTs and DCFTs. Our proposal reduces to the known ones in the shaded regions and provides new ones in the region colored in light blue.}
\label{table1}
\end{table}

While we demonstrate a few field theoretic examples as supporting evidences for our conjecture, we are able to provide a  holographic proof under the assumption of the null energy condition in several holographic models of defect RG flows.
We suspect our conjecture may be proven at least for $p=1$ by suitably extending the argument of \cite{Casini:2016fgb} for the entropic proof of the $g$-theorem.

The organization of this paper is as follows.
In section \ref{sec2}, we review the CHM map in DCFTs and discuss the structures of and the relation between the defect free energy and the defect entropy.
In section \ref{sec3}, we propose to use the universal part of the defect free energy as a $C$-function.
We then test our proposal with several examples of DCFTs.
In section \ref{sec4}, we consider various holographic models of DCFTs and give a holographic proof of our conjecture.
Finally section \ref{con} is devoted to conclusion and discussion.
Appendix \ref{notation} summarizes our notation.
In appendix \ref{ap:Relative} we consider the relative entropy between DCFT and CFT as another measure and show the equivalence to the defect free energy.

\section{Sphere partition function and entanglement entropy in DCFT} \label{sec2}
In this section we consider two quantities: the defect free energy and the defect entropy.
The former is the increment of the sphere free energy from a conformal defect in DCFT while the latter is the additional contribution to the entanglement entropy of a spherical region.
To set the stage, we begin with reviewing the implication of defects for the correlation functions in DCFT.
We then turn to describing the conformal transformation known as the CHM map which relates the entanglement entropy of a spherical region to the thermal entropy of DCFT on a hyperbolic space.
With this relation we derive a formula expressing the defect entropy by the defect free energy and the one-point function of the stress-energy tensor.
The structure of the UV divergent terms is also discussed for the defect entropy and the defect free energy to identify their universal parts independent of the regularization scheme.

\subsection{Defect CFT}
Defects collectively stand for non-local operators in QFT as exemplified by Wilson-'t Hooft line operators.
A certain class of defects has realizations by fundamental fields in a given QFT (e.g. Wilson lines) while some are rather defined by specifying boundary conditions around them on the fundamental fields (e.g. 't Hooft lines).
One can also couple a lower-dimensional theory to a higher-dimensional theory  (e.g. the mixed-dimensional QED and the D3/D5 brane model).
Thus there are at least three different ways to introduce defects\footnote{These constructions may be equivalent in certain cases while we are not aware of their precise relations.} \cite{Gaiotto:2014ina}:
\begin{enumerate}
\item Localize the ambient fields at the location of the defect. 
\item Impose a boundary condition on the ambient fields around the defect 
\cite{Kapustin:2005py,Gukov:2006jk}.
\item Introduce new degrees of freedom localized on the defect and couple them to the ambient fields.
\end{enumerate}
When $p=d-1$, we can instead introduce a boundary or an interface by gluing two different theories along a boundary.

In this paper we restrict our attention to a special class of defects, called conformal defects, which are hyperplaner or spherical to preserve the conformal symmetry on and the rotational symmetry around the worldvolumes. 
Conformal defects of dimension $p$ break the ambient conformal symmetry $\SO(1,d+1)$ to the subgroup $\SO(1,p+1) \times \SO(d-p)$, which turns out to be strong enough to constrain the correlation functions in defect CFT.

While the correlation functions of defect local operators are determined by the same argument as in CFT, there are other class of correlation functions involving the ambient operators in DCFT which can still be fixed by the residual conformal symmetry \cite{McAvity,Billo:2016cpy}.
In particular, the one-point function of an ambient operator does not necessarily vanish in DCFT.

To illustrate this point in detail, let us consider a ($p$-dimensional) planer defect in $\BR^d$ and the stress-energy tensor.
The metric is then divided into the parallel and orthogonal components:
\begin{align}
	\dd s^2 = \dd \hat{x}^a\, \dd \hat{x}^a + \dd x_\perp ^i\, \dd x_\perp ^i\ ,  \qquad (a = 0, \cdots ,\,p-1, \ i = p, \cdots ,\,d-1) \ .
\end{align}
First, we deal with the cases 1 and 3 in the aforementioned classification. 
Assuming DCFT has a Lagrangian description, the Lagrangian consists of the ambient part and the defect part, 
\begin{align}
	I_{\text{DCFT}} = \int \dd ^dx\sqrt{g}\,\CL_{\text{CFT}} + \int \dd^p \hat{x} \sqrt{\hat{g}}\,\hat{\CL}_{\text{defect}} \ .
\end{align}
In the case 1, the defect part is absent, but a defect operator $\CD^{(p)}$ should be inserted in evaluating correlation functions \cite{Billo:2016cpy}
\begin{align}
\langle\, \mathcal{O}\cdots  \mathcal{O} \,\rangle_{\mathcal{D}^{(p)}} \equiv \frac{\langle\, \mathcal{O}\cdots  \mathcal{O}\, \mathcal{D}^{(p)}\, \rangle}{\langle\, \mathcal{D}^{(p)} \,\rangle} \ .
\end{align}
We are then allowed to regard $-\log\,\mathcal{D}^{(p)}$ as the defect part in the action.
In either case the stress-energy tensor follows from the partition function $Z^\text{(DCFT)}$\footnote{The definition differs in the sign from the one used in \cite{Billo:2016cpy}, so $T_\text{DCFT}$ equals $-T_\text{tot}$ there.}
\begin{align}\label{stress-tensor-def}
	T^{\mu\nu}_\text{DCFT} = -\frac{2}{\sqrt{g}}\, \frac{\delta\, \log Z^\text{(DCFT)}[g_{\mu\nu}]}{\delta g_{\mu\nu}} \ .
\end{align}
It will be useful to split it into the ambient part $T^{\mu\nu}_\text{CFT}$ and the defect localized part $t^{\mu\nu}$
\begin{align}
	T_\text{DCFT}^{\mu\nu} = T^{\mu\nu}_\text{CFT} + t^{\mu\nu}\ .
\end{align}
$t^{\mu\nu}$ contains the contribution from the response to the induced metric \cite{Billo:2016cpy}
\begin{align}
	t^{\mu\nu} = \delta_\CD (x_\perp) \left[\delta_a^\mu\delta_b^\nu\, B^{ab} + \cdots \right] + \frac{1}{2}\,\partial_i\delta_\CD (x_\perp)\,\delta_a^\mu\delta_b^\nu\,C^{abi} +\cdots  \ ,
\end{align}
where $\delta_\CD (x_\perp)$ is the delta function localized on the worldvolume of the defect and $B^{ab}$ and $C^{abi}$ are defined by the variation of the defect action (see \cite{McAvity,Billo:2016cpy,Armas:2017pvj} for the detail).
In what follows, we ignore the higher derivative terms of the delta function as they vanish for the planar defect.
While the conservation and tracelessness of the ambient stress tensor are violated in the presence of the defect, $T_\text{DCFT}^{\mu\nu}$ is traceless and partially conserved
\begin{align}
	\begin{aligned}
		\partial_\mu T^{\mu a}_\text{DCFT} &= 0 \ , \\
        \partial_\mu T^{\mu i}_\text{DCFT} &= -\delta_\CD (x_\perp) \, \mathrm{D}^i 
        \ , \\
        (T_\text{DCFT})^\mu_{~\mu} &= 0 \ .
	\end{aligned}
    \label{emtensor}
\end{align}
These relations hold as the operator identities in DCFT.

In contrast to the cases 1 and 3, the Lagrangian and the stress-energy tensor in the case 2 are the same as those in CFTs without a defect.
However, a careful treatment is required in evaluating the one-point function of the stress-energy tensor as we will discuss later on.

Now consider the one-point function of the ambient stress-energy tensor, $T_\text{CFT}$. $T_\text{CFT}$ is a symmetric traceless tensor of dimension $d$ and spin 2, hence the residual conformal symmetry completely fixes the form of the correlator
\begin{align}\label{ST_one_point}
	\begin{aligned}
	\langle\, T^{ab}_{\text{CFT}}(x) \,\rangle &=   \frac{d-p-1}{d} \frac{a_T}{|x_\perp|^d} \,\delta^{ab}\ , \\
    \langle\, T^{ij}_{\text{CFT}}(x)\, \rangle &= - \frac{a_T}{|x_\perp|^d} \left(\frac{p+1}{d}\delta^{ij} -  \frac{x_\perp^i x_\perp^j}{|x_\perp|^2}  \right)\ , \\
    \langle\, T^{ai}_{\text{CFT}} (x) \, \rangle &= 0 \ ,
    \end{aligned}
\end{align}
where $a_T$ is a constant characterizing the defect.\footnote{Our $a_T$ is $-a_\CT$ in \cite{Billo:2016cpy}.}
While the one-point function does not vanish in general, $\langle \,T_{\text{CFT}}^{\mu\nu}\, \rangle = 0$ for $p=d-1$, including interface CFT and BCFT, as seen from \eqref{ST_one_point}.
More generically, the one-point function of the ambient operator with non-zero spin vanishes in BCFT and DCFT with a defect of dimension $d-1$ \cite{McAvity,Liendo:2012hy,Fukuda:2017cup}.

Furthermore, the one-point function of $t^{\mu \nu}$ vanishes 
\begin{align}\label{Defect_ST_vanish}
\langle\, t^{\mu \nu}(x)\, \rangle =0 \ .
\end{align}
This is seen by writing $t^{\mu \nu}$ as
\begin{align}
t^{\mu \nu}(x) =\delta_\CD (x_\perp)\, \frac{\partial x^\mu}{\partial \hat{x}^a} \frac{\partial x^\nu}{\partial \hat{x}^b}\, \hat{t}^{ab} (\hat{x})\ ,
\end{align}
and define the defect stress-energy tensor $\hat{t}^{ab} (\hat{x})$, which is a defect local operator of dimension $p$ whose vev must be zero due to the invariance under the translation, rotation and scale transformation on the defect.

When $p$ is even there exist conformal anomalies (the Graham-Witten anomaly \cite{Graham:1999pm}), but we assume the dimensional regularization for both $d$ and $p$ so as to avoid them in the rest of this paper.

\subsection{CHM map}
\begin{figure}
\begin{tabular}{ccc}
  \begin{minipage}{0.35\hsize}
  \centering
  \begin{tikzpicture}[scale=1.2]
   \draw (0.5,0.4)--(3.9,0.4)--(5.5,1.6)--(2.1,1.6)--cycle;
   \draw[-{Triangle[angle'=60,scale=0.8]}] (2.1, 1.6)--(2.1, 2.5) node [right] {$t$};
   \draw (3.0,1.0)--(3.0,2.5) node [right] {$\mathcal{D}^{(1)}$};
   \draw[dashed] (3.0,0.4)--(3.0,1.0);
   \draw (3.0,-0.4)--(3.0,0.4);
   \draw[{Triangle[angle'=60,scale=0.8]}-{Triangle[angle'=60,scale=0.8]}] (2.0,0.7)--(2.4,0.85) node [above] {$R$} --(3.0,1.0);
   \draw[orange, thick] (4.25,1) arc [start angle = 0, end angle = 180, x radius=1.25cm, y radius=0.5cm];
   \draw[{Triangle[angle'=60,scale=0.6]}-] (4.2,1.2)--(4.5,1.3) node [right] {$\Sigma$};
   \draw[orange, thick] (4.25,1) arc [start angle = 0, end angle = -180, x radius=1.25cm, y radius=0.5cm];
   \draw (1,1.7) node [above] {$\mathbb{R}^{1,d-1}$};
   \draw (3.5,0.7) node [above] {$A$};
  \end{tikzpicture}
\end{minipage}
\begin{minipage}{0.35\hsize}
  \centering
   \begin{tikzpicture}[scale=1.2]
   \draw (0.5,0.4)--(3.9,0.4)--(5.5,1.6)--(2.1,1.6)--cycle;
   \draw[-{Triangle[angle'=60,scale=0.8]}] (2.1, 1.6)--(2.1, 2.5) node [right] {$t$};
   \draw[dashed] (2.2,0.4)--(3.8,1.6);
   \draw(3.8,2.4) node [right] {$\mathcal{D}^{(d-1)}$};
   \filldraw[fill=gray,opacity=0.3,draw=none] (2.2,-0.4)--(3.8,0.8)--(3.8,2.4)--(2.2,1.2)--cycle;
   \draw (3.266,0.4)--(2.2,-0.4)--(2.2,1.2)--(3.8,2.4);
   \draw (3.8,1.6)--(3.8,2.4);
   \draw[dashed,opacity=0.5] (3.266,0.4)--(3.8,0.8)--(3.8,1.6);
   \draw[orange, thick] (4.25,1) arc [start angle = 0, end angle = 180, x radius=1.25cm, y radius=0.5cm];
   \draw[orange, thick] (4.25,1) arc [start angle = 0, end angle = -180, x radius=1.25cm, y radius=0.5cm];
   \end{tikzpicture}
\end{minipage}
\begin{minipage}{0.3\hsize} 
  \centering
  \begin{tikzpicture}[scale=1.2]
   \draw (0.2,0.4)--(2.0,0.4)--(3.6,1.6)--(1.8,1.6)--cycle;
   \draw[-{Triangle[angle'=60,scale=0.8]}] (1.8, 1.6)--(1.8, 2.5) node [left] {$t$};
   \draw(1.9,2.4) node [right] {$\mathcal{D}^{(d-1)}$};
   \filldraw[fill=gray,opacity=0.3,draw=none] (0.2,-0.4)--(1.8,0.8)--(1.8,2.4)--(0.2,1.2)--cycle;
   \draw (1.266,0.4)--(0.2,-0.4)--(0.2,1.2)--(1.8,2.4);
   \draw (1.8,1.6)--(1.8,2.4);
   \draw[dashed,opacity=0.5] (1.266,0.4)--(1.8,0.8)--(1.8,1.6);
   \draw[orange, thick] (0.45,0.585) arc [start angle = -118, end angle = 62, x radius=1.25cm, y radius=0.5cm];
  \end{tikzpicture}
\end{minipage}
\end{tabular}
\caption{(Left) A dimension-one conformal defect $\mathcal{D}^{(1)}$ in  Lorentzian flat spacetime. The spherical subsystem $A$ of radius $R$ surrounds the defect. (Center, Right) A codimension-one defects $\mathcal{D}^{(d-1)}$ as an interface (Center) and a boundary (Right). The subsystem $A$ intersects with the defect in these cases.}
\label{fig:setup}
\end{figure}
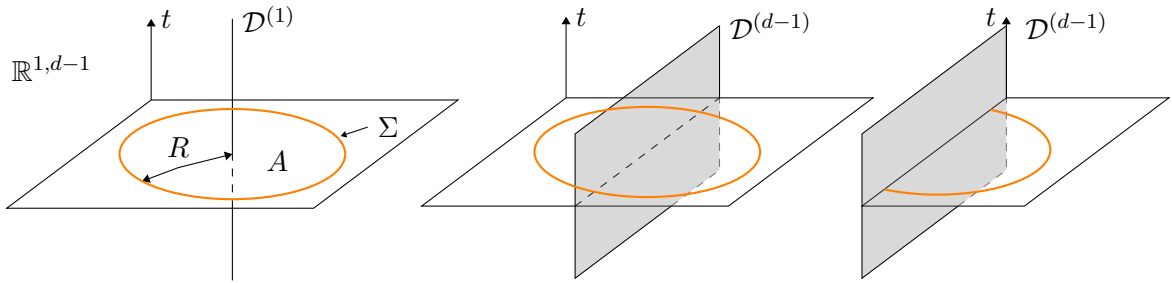

In the rest of this section, we are concerned with the entanglement entropy across a sphere in defect CFT$_d$.
To this end, it is convenient to adopt the polar coordinates of the flat space $\BR^{1,d-1}$ in Lorentzian signature,
\begin{align}
\begin{aligned}
	\d s_{\BR^{1,d-1}}^2 &= \eta_{\mu\nu}\, \d X^\mu\, \d X^\nu  \\
			&= -\d t^2 + \d r^2 + r^2\,\d s_{\mathbb{S}^{d-2}}^2 \ ,
\end{aligned}
\end{align}
where $\eta_{\mu\nu} = \text{diag} (-, +, \cdots, +)$ and the entangling surface $\Sigma$ is a $(d-2)$-dimensional hypersphere of radius $R$ located at $t=0$ time slice:
\begin{align}
	\Sigma = \{ X^0 = t=0,\, r= R\} \ .
\end{align}
We want to introduce a \emph{conformal defect} $\CD^{(p)}$ of dimension-$p$ respecting the subgroup $\SO(2, p) \times \SO (d-p)$ of the conformal group $\SO(2,d)$.
Conformal defects are either planer or spherical, and we choose $\CD^{(p)}$ to be a hyperplane,
\begin{align}
	\CD^{(p)} = \{ X^{p} = \cdots =  X^{d-1} = 0 \} \ .
\end{align}
Figure \ref{fig:setup} shows our setups for $p=1$ and $p=d-1$.

Using the replica trick, the entanglement entropy across $\Sigma$ amounts to the calculation of the partition function on the branched cover $\CM_n$ of $\BR^{1,d-1}$.
It is most easily performed with the inverse transformation of the CHM map
\cite{Casini:2011kv,Jensen:2013lxa},\footnote{We will focus on the case with $1 \leq p < d-1$ so that this map works, but the following results hold for $p=d-1$ with a slight change \cite{Jensen:2013lxa}.}
\begin{align}
	x^\mu (X) = 4\left[ \frac{X^\mu - |X|^2\, C^\mu  }{1 - 2 X\cdot C + |X|^2\, |C|^2} + \frac{R^2}{2}\, C^\mu\right] \ , \qquad C^\mu\partial_\mu = - \frac{1}{R}\,\partial_1 \ .
\end{align}
The resulting space is conformally flat with the metric,
\begin{align}
	\d s_{\BR^{1,d-1}}^2 &= \Omega(x)^2\,\eta_{\mu\nu}\,\d x^\mu\, \d x^\nu \ , 
\end{align}
with the conformal factor,
\begin{align}
	\begin{aligned}
	\Omega &= \frac{1}{4}\,(1 - 2 X\cdot C + |X|^2\, |C|^2 ) \\
		&= \frac{1}{1 + x\cdot C + |x|^2\, |C|^2/4} \ .
	\end{aligned}
\end{align}
After the conformal transformation, the causal domain $r \pm t \le R$ for the entangling region maps to the (right) Rindler wedge $x^\pm\equiv x^1 \pm x^0 \ge 0$,
and the light cones $r+ t = R$ and $r- t = R$ on the boundary of the causal domain are mapped to the Rindler horizons,
\begin{align}
	\begin{aligned}
		r+ t &= R \quad \Rightarrow \quad x^+ = 0 \ , \\
		r- t &= R \quad \Rightarrow \quad x^- = 0 \ ,
	\end{aligned}
\end{align}
The entangling surface is mapped to the origin in the $x^0$-$x^1$ plane,
\begin{align}
	\Sigma = \{ x^0 = x^1 = 0 \} \ ,
\end{align}
while the defect is mapped to the hyperplane,
\begin{align}
	\CD^{(p)} = \{ x^{p} = \cdots = x^{d-1} = 0 \} \ .
\end{align}

\paragraph{Hyperbolic coordinates}
Introducing the new coordinates,
\begin{align}
	x^\pm = z\, \mathrm{e}^{\pm \tau} \ ,
\end{align}
the Rindler space becomes
\begin{align}
	\begin{aligned}
	\d s^2_\text{Rindler} &= \d x^+\, \d x^- + \sum_{i=2}^{d-1}(\d x^i)^2 \\
		&= z^2\left[ - \d \tau^2 + \frac{\d z^2 + \sum_{i=2}^{d-1}(\d x^i)^2}{z^2}\right] \ ,
	\end{aligned}
\end{align}
which is conformally equivalent to $\BR \times \BH^{d-1}$ parametrized by $\tau$ and 
a hyperbolic space of unit radius,
\begin{align}
	- y_{0}^2 + y_1^2 + y_2^2 + \cdots + y_{d-1}^2 = -1 \ ,
\end{align}
in the Poincar\'e coordinates,
\begin{align}
	\begin{aligned}
		y_0 &= \frac{z}{2}\left[ 1 + \frac{1 + \sum_{i=2}^{d-1}(x^i)^2}{z^2} \right]\ , \\
		y_1 &= \frac{z}{2}\left[ 1 + \frac{-1 + \sum_{i=2}^{d-1}(x^i)^2}{z^2} \right] \ , \\
		y_i  &= \frac{x^i}{z} \ , \qquad (i=2, \cdots, d-1) \ .
	\end{aligned}
\end{align}
In these new coordinates, the entangling surface and the defect are located at\footnote{The position of $\Sigma$ in the $\tau$ direction is ambiguous as the $\tau$ circle shrinks at $z=0$. We thus choose a reference point at $\tau=0$.}
\begin{align}
	\Sigma = \{ z=0 \ , \tau = 0 \} \ , \qquad \CD^{(p)} = \{ x^{p} = \cdots = x^{d-1} = 0 \} \ .
\end{align}

For later convenience, we introduce the global coordinates of $\BH^{d-1}$ by
\begin{align}
	\begin{aligned}
		y_a &= \cosh x \, f_a\ , &\qquad &(a=0, \cdots, p-1)\ , \\
		y_i  &= \sinh x\, e_i \ , &\qquad &(i=p, \cdots, d-1) \ .
	\end{aligned}
\end{align}
where $-f_0^2 + \sum_{a=1}^{p-1}\,f_a^2 = -1$ and $\sum_{i=p}^{d-1}\, e_i^2 = 1$.
The resulting metric for $\BR \times \BH^{d-1}$ takes the form
\begin{align}\label{Hyperbolic}
	\d s_{\BR \times \BH^{d-1}}^2 = -\d \tau^2 + \d x^2 + \cosh^2 x\, \d s^2_{\BH^{p-1}} + \sinh^2 x \, \d s_{\mathbb{S}^{d-p-1}}^2 \ ,
\end{align}
where the entangling surface and the defect are situated at
\begin{align}
	\Sigma = \{ x=\infty \ , \tau = 0 \} \ , \qquad \CD^{(p)} = \{ x = 0 \} \ .
\end{align}

\paragraph{de Sitter $\times$ hyperbolic coordinates}
We will make one more coordinate transformation $\sinh x = \cot \theta$ which takes us from \eqref{Hyperbolic} to another coordinate system, 
\begin{align}
	\d s_{\BR \times \BH^{d-1}}^2 = \frac{1}{\sin^2\theta}\, \d s_{\text{dS}_{d-p+1}\times \BH^{p-1}}^2 \ ,
\end{align}
where the static patch of the de Sitter space is employed,
\begin{align}\label{dS}
	\d s_{\text{dS}_{d-p+1}\times \BH^{p-1}}^2 =  -\sin^2\theta\, \d \tau^2  + \d \theta^2 + \cos^2\theta\, \d s_{\mathbb{S}^{d-p-1}}^2 + \d s^2_{\BH^{p-1}} \ ,
\end{align}
with $0\le \theta \le \pi/2$.
The entangling surface and the defect are mapped to
\begin{align}
	\Sigma = \{ \theta=0 \ , \tau = 0 \} \ , \qquad \CD^{(p)} = \{ \theta = \pi/2 \} \ .
\end{align}

\subsection{Sphere partition function and defect entropy}
We have shown the flat spacetime is conformally equivalent to both $\BR\times \BH^{d-1}$ in \eqref{Hyperbolic} and the de Sitter $\times$ hyperbolic space in the static patch \eqref{dS}.

In Euclidean signature, the former becomes 
\begin{align}\label{EHyper}
	\d s_{\BS^1 \times \BH^{d-1}}^2 = \d \tau^2 + \d x^2 + \cosh^2 x\, \d s^2_{\BH^{p-1}} + \sinh^2 x \, \d s_{\mathbb{S}^{d-p-1}}^2 \ ,
\end{align}
by Wick rotation $\tau \to \i \,\tau$ while the de Sitter subspace in \eqref{dS} becomes a sphere in the latter case,
\begin{align}\label{Sphere}
	\d s_{\BS^{d-p+1}\times \BH^{p-1}}^2 = \d \theta^2 + \sin^2\theta\, \d \tau^2 +\cos^2\theta\, \d s_{\mathbb{S}^{d-p-1}}^2 + \d s^2_{\BH^{p-1}} \ ,
\end{align}
with the ranges $0\le \theta \le \pi/2$ and $0\le \tau< 2\pi$.

\begin{figure}
\centering
     \begin{tikzpicture}[scale=1.5]
    \draw (0,0) parabola (-4,1);
    \draw (0,0) parabola (-4,-1);
    \draw (-1,0) circle [x radius=0.4mm, y radius=0.65mm];
    \draw (-2,0) circle [x radius=0.75mm, y radius=2.5mm];
    \filldraw[pattern=horizontal lines, pattern color=gray] (-3,0) circle [x radius=1mm, y radius=5.5mm];
    \draw (-4,0) circle [x radius=1.25mm, y radius=10mm];
    \draw[-{Triangle[angle'=60,scale=0.8]}] (-1.5,0.4) parabola (-2.7,0.8) node [above] {$x$};
    \draw[{Triangle[angle'=60,scale=0.8]}-] (-1.85,-0.55) node [right] {$\tau$} arc [start angle = -60, end angle = 140, x radius = 2mm, y radius = 6mm];
    \draw (0,0) node [below right] {$x=0$};
    \draw (0,0) node [above right] {$\mathcal{D}^{(p)}$};
    \draw (-4.1,1) node [left] {$\Sigma$};
    \draw (-4.3,1.2) node [above] {$\tau=0, x=\infty$};
    \fill (0,0) circle [radius=0.6mm];
    \fill (-4,1) circle [radius=0.6mm];
    \draw[{Triangle[angle'=60,scale=0.8]}-] (-3.1,-0.1) arc [start angle = 100, end angle = 260, x radius = 2.5mm, y radius = 4mm] node [right] {$\mathbb{S}^{d-p-1}$};
    \draw[dashed] (-1.2,0)--(-1.2,0.9);
    \draw (-0.8,1.3) node [above] {$\mathbb{H}^{p-1}$};
    \draw (-1.2,0) circle [radius=0.3mm];
    \draw (-1.0+0.5,1.1+0.3) parabola bend (-1.0,1.1+0.1) (-1.0-0.5,1.1+0.3);
    \draw (-1.0+0.5,1.1-0.3) parabola bend (-1.0,1.1-0.1) (-1.0-0.5,1.1-0.3);
    \draw (-1.0,1.1) circle [x radius=0.5mm, y radius=1mm];
    \draw (-1.5,1.1) circle [x radius=1mm, y radius=3mm];
    \draw (-0.5,0.8) arc [start angle=-90, end angle=90 ,x radius=1mm, y radius=3mm];
    \draw[dashed] (-0.5,1.4) arc [start angle=90, end angle=270 ,x radius=1mm, y radius=3mm];
    \end{tikzpicture}
    \caption{The locations of the entanglement surface $\Sigma$ and the conformal defect $\mathcal{D}^{(p)}$ in the hyperbolic coordinates \eqref{EHyper}. The hyperbolic space $\mathbb{H}^{p-1}$ is fibered on each point of the base.}
    \label{fig:Hyperbolic}
\end{figure}
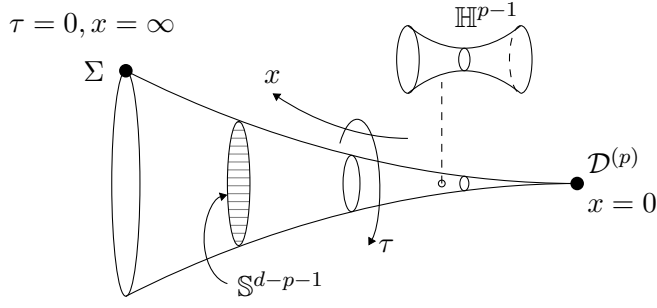

The entangling surface and the defect are located at
\begin{align}
	\Sigma = \{ x=\infty \ , \tau = 0 \} \ , \qquad \CD^{(p)} = \{ x = 0 \} \ ,
\end{align}
in the former (see figure \ref{fig:Hyperbolic}) and 
\begin{align}
	\Sigma = \{ \theta=0 \ , \tau = 0 \} \ , \qquad \CD^{(p)} = \{ \theta = \pi/2 \} \ ,
\end{align}
in the latter.
It is obvious from \eqref{Sphere} that the defect located at $\theta = \pi/2$ is a subspace $\BS^1 \times \BH^{p-1}$.

Recalling the definition of the R{\'e}nyi entropy in the replica trick,
\begin{align}
	S_n = \frac{1}{1-n}\, \log \frac{Z[\CM_n]}{\left( Z[\CM_1]\right)^n} \ ,
\end{align}
the calculation of entanglement entropy ends up with knowing the partition function $Z[\CM_n]$ on the $n$-fold cover $\CM_n$.
For a spherical entangling region in CFT, the $n$-fold cover is conformally equivalent to the $n$-fold cover $\BS^{1}_n \times \BH^{d-1}$ along the $\tau$ coordinate of the space \eqref{EHyper},
\begin{align}\label{Branched_Sphere}
		\d s_{\BS^{1}_n \times \BH^{d-1}}^2 &=  n^2\, \d \tau^2 + \d x^2 + \cosh^2 x\, \d s^2_{\BH^{p-1}} + \sinh^2 x \, \d s_{\mathbb{S}^{d-p-1}}^2 \ ,
\end{align}
with the range $0\le \tau < 2\pi$.
If there are no conformal anomalies, the partition function is 
invariant under the conformal map,
\begin{align}
	Z[\CM_n] = Z [\BS^{1}_n \times \BH^{d-1}] \ .
\end{align}
Hence the R{\'e}nyi entropy across a sphere in CFT is given by,
\begin{align}\label{SphereRE}
	S_n^\text{(CFT)} = \frac{1}{1-n}\, \log \frac{Z^\text{(CFT)}[\BS^{1}_n \times \BH^{d-1}]}{\left( Z^\text{(CFT)}[\BS^{1} \times \BH^{d-1}]\right)^n} \ .
\end{align}

Now we consider a defect CFT with a conformal defect $\CD^{(p)}$ of dimension-$p$ for $p\le d-2$ and defer the discussion for $p=d-1$ to the end of this subsection.
We then define the \emph{defect entropy} by the additional entanglement entropy due to the existence of $\CD^{(p)}$:
\begin{align}
	S_\text{defect} \equiv \lim_{n\to 1}\, \left( S_{n}^\text{(DCFT)} - S_{n}^\text{(CFT)}\right) \ .
\end{align}
The R\'enyi entropy $S_{n}^\text{(DCFT)}$ in DCFT is defined in a similar manner to the R\'enyi entropy in CFT, which takes the same form as \eqref{SphereRE} for a spherical entangling region:
\begin{align}\label{DCFT_RE}
	S_{n}^\text{(DCFT)} = \frac{1}{1-n}\,  \log \frac{Z^\text{(DCFT)}[\BS^{1}_n \times \BH^{d-1}]}{\left( Z^\text{(DCFT)}[\BS^{1} \times \BH^{d-1}]\right)^n}\ .
\end{align}
Hence it is more efficient to rewrite the defect entropy as
\begin{align}
	S_\text{defect} \equiv \lim_{n\to 1}\, \frac{1}{1-n}\,  \log\, \frac{\langle\, \CD^{(p)}\,\rangle_n}{\langle\, \CD^{(p)}\, \rangle^n}\ ,
\end{align}
where $\langle\, \CD^{(p)}\,\rangle_n$ is the vev of the conformal defect operator $\CD^{(p)}$ of dimension $p$ on $\BS^d$,
\begin{align}\label{I_vev}
	\langle\, \CD^{(p)}\,\rangle_n \equiv \frac{Z^\text{(DCFT)}[\BS^{1}_n \times \BH^{d-1}]}{Z^\text{(CFT)}[\BS^{1}_n \times \BH^{d-1}]} \ .
\end{align}
We also denote $\langle\, \CD^{(p)}\, \rangle \equiv \langle\, \CD^{(p)}\, \rangle_1$ to simplify the notation.

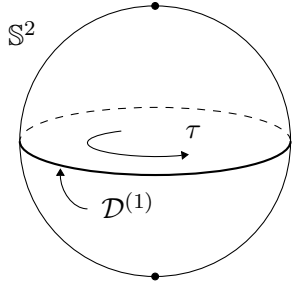
\begin{figure}
   \centering
   \begin{tikzpicture}[scale=1.8]
    \draw (0,0) circle [radius = 1cm];
    \draw[dashed] (1,0) arc [start angle=0, end angle=180, x radius = 1cm, y radius= 0.25cm];
    \draw[thick] (1,0) arc [start angle=0, end angle=-180, x radius = 1cm, y radius= 0.25cm];
    \draw[-{Triangle[angle'=60,scale=0.8]}] (-0.25,0.075) arc [start angle=120, end angle=300, x radius = 5mm, y radius= 1mm];
    \draw (-0.8,0.8) node [left] {$\mathbb{S}^{2}$};
    \draw (-0.2,-0.3) node [below] {$\mathcal{D}^{(1)}$};
    \draw[-{Triangle[angle'=60,scale=0.8]}] (-0.5,-0.5) arc [start angle=-90, end angle=-180, x radius=2.0mm, y radius=3.0mm];
    \filldraw[black] (0,1) circle [radius=0.25mm];
    \filldraw[black] (0,-1) circle [radius=0.25mm];
    \draw (0.15,-0.05) node [above right] {$\tau$};
    \end{tikzpicture}
    \caption{The conformal defect $\mathcal{D}^{(1)}$ on $\mathbb{S}^{d}$. For $d=2$, it winds along the equator of $\mathbb{S}^{2}$ ($\tau$-direction).}
    \label{fig:sphere}
\end{figure}

Note that for CFT the sphere entanglement entropy equals to the partition function on a conformally flat space up to UV divergences \cite{Casini:2011kv},\footnote{See, however, \cite{Rodriguez-Gomez:2017kxf} where a variant of conformal anomalies was observed even in odd $d$ dimensions.}
\begin{align}
	S^\text{(CFT)} = \log\, Z^\text{(CFT)}[\BS^d]  = \log\, Z^\text{(CFT)}[\BS^{1} \times \BH^{d-1}]\ .
\end{align}
In order to derive a similar relation for the defect entropy, 
we expand the partition function $Z^\text{(DCFT)}[\BS^{1}_n \times \BH^{d-1}]$ on the branched space \eqref{Branched_Sphere} around $n=1$,
\begin{align}
	\begin{aligned}
	\log\, Z^\text{(DCFT)}[\BS^{1}_n \times \BH^{d-1}] &= \log\, Z^\text{(DCFT)}[\BS^{1} \times \BH^{d-1}] \\
		&\qquad - \frac{1}{2}\int_{\BS^{1} \times \BH^{d-1}}\, \delta g_{\tau\tau}\, \langle\, (T_\text{DCFT})^{\tau \tau}\,\rangle_{\BS^{1} \times \BH^{d-1}}^\text{(DCFT)} + \cdots \ ,
	\end{aligned}
\end{align}
where $\delta g_{\tau\tau} = (n^2 -1)$. 
Since $\cdots$ are terms of order $(n-1)^2$ which do not contribute to the entanglement entropy we obtain
\begin{align}\label{Defect_Entropy_bare}
	S_\text{defect} = \log\, \langle\, \CD^{(p)}\, \rangle + \int_{\BS^{1} \times \BH^{d-1}}\, \langle\, (T_\text{DCFT})^\tau_{~\tau}\,\rangle_{\BS^{1} \times \BH^{d-1}}^\text{(DCFT)} \ .
\end{align}
We will call the first term in the right hand side the \emph{defect free energy}, which can be written by the sphere free energy,
\begin{align}\label{Defect_Free_Energy}
	\log\, \langle\, \CD^{(p)}\,\rangle = \log \, \frac{Z^\text{(DCFT)}[\BS^d]}{Z^\text{(CFT)}[\BS^d]} \ ,
\end{align}
if there are no conformal anomalies.
After the conformal transformation, the defect which was originally planer on flat space is mapped to a spherical defect on $\BS^d$ (see figure \ref{fig:sphere}).

When located on a point away from the defect the one-point function of the stress-energy tensor in DCFT is fixed on $\BS^{1} \times \BH^{d-1}$ by imposing the conformal symmetry, tracelessness and conservation law to the following form:
\begin{align}\label{Stress_tensor_One_Point}
	\begin{aligned}
	&\langle\, (T_\text{DCFT})_{\mu\nu}\, \rangle_{\BS^{1} \times \BH^{d-1}}^\text{(DCFT)}\, \d x^\mu \otimes \d x^\nu \\
    &\qquad = \frac{a_T}{\sinh^d x} \left[ \frac{d-p-1}{d} \left(  \d \tau^2 + \d x^2 + \cosh^2 x\, \d s^2_{\BH^{p-1}} \right) - \frac{p+1}{d} \sinh^2 x \, \d s_{\mathbb{S}^{d-p-1}}^2 \right] \ ,
    \end{aligned}
\end{align}
where $a_T$ is the same as the one in \eqref{ST_one_point}.
In principle the defect localized term could appear in \eqref{Stress_tensor_One_Point}, but such a term vanishes due to \eqref{Defect_ST_vanish} when DCFT has a Lagrangian description.
If the defect is defined as a boundary condition for the ambient fields, there is no defect localized part in the stress tensor, but the boundary condition still affects the ambient stress tensor in the same way as \eqref{Stress_tensor_One_Point}.

In evaluating the one-point function of the stress tensor integrated over $\BS^{1} \times \BH^{d-1}$ in \eqref{Defect_Entropy_bare}, one needs a regularization for the UV divergence arising from the integration near the defect.
We follow the prescription employed by \cite{Kapustin:2005py,Lewkowycz:2013laa} and remove the tubular neighborhood of a defect in flat space whose boundary is $\BR^p \times \BS^{d-p-1}$, on which we impose a boundary condition for the ambient fields.
After performing the CHM map, this regularization amounts to restricting the integration range of $x$ to $\epsilon \le x < \infty$ for a small parameter $\epsilon$ and sending $\epsilon\to 0$ in the end.
This prescription makes it manifest that there are no contributions from the defect localized term in the stress tensor.
Expanding in the small $\epsilon$ one can read off the constant part of the integral, but a more illuminating way is to use the dimensional regularization in $d$ after setting $\epsilon =0$.
The two methods agree on giving the same universal constant.

In either way one can evaluate the integral in \eqref{Defect_Entropy_bare} with only the ambient term of the stress tensor \eqref{Stress_tensor_One_Point} and derive the universal formula for the defect entropy:

\bigskip
{\it
	In a DCFT with a conformal defect of dimension $p\le d- 2$ the defect entropy of a spherical entangling surface is given, up to UV divergence, by
\begin{align}\label{Defect_Entropy_universal}
	S_\mathrm{defect} = \log\, \langle\, \CD^{(p)}\,\rangle - \frac{2(d-p-1)\,\pi^{d/2+1} }{\sin\left(\pi p/2\right)\,d \,\Gamma\left(p/2+1\right)\,\Gamma \left( (d-p)/2\right)}\, a_T \ .
\end{align}
}

\noindent
This is one of our main results.
This formula is seen as a generalization of the result for $p=1$ \cite{Lewkowycz:2013laa}.
For clarifying the validity of this formula, a few comments are in order:
\begin{itemize}
	\item In deriving \eqref{Defect_Entropy_universal}, we assume 
		\begin{enumerate}
		\item[(i)] there are no conformal anomalies, i.e., we consider DCFTs in continuous dimensions, 
		\item[(ii)] the $n$-dependence is only through the metric,
		\item[(iii)] the metric is coupled to the conformal stress-energy tensor.
		\end{enumerate}
		The last two assumptions should be regarded as the ``choice" of the R\'enyi entropy in QFTs, and may vary depending on the situation.
        For instance, one can choose a boundary condition around the entangling surface so as to respect supersymmetry \cite{Nishioka:2013haa}.
        Then the $n$-dependence is not only through the metric, but also arises from the background fields of supergravity.
	\item The one-point function of the stress tensor in \eqref{Stress_tensor_One_Point} is renormalized and the identity holds up to UV divergences that can be removed by counterterms to a background gravitational theory.
		Hence \eqref{Defect_Entropy_universal} holds only up to UV divergences.
	\item There are Graham-Witten type conformal anomalies \cite{Graham:1999pm} for $p$ even, which is manifest in \eqref{Defect_Entropy_universal} as a pole of the sine function and produces the logarithmic divergence.
\end{itemize}

In the case with $p=d-1$ there are two types of theories, BCFTs and the others, depending on whether they are defined on a manifold with boundary or not.
For BCFTs we should define the \emph{boundary entropy} by
\begin{align}\label{Boundary_Entropy}
	S_\text{bdy} \equiv \lim_{n\to 1}\, \left( S_{n}^\text{(BCFT)} - \frac{1}{2}\,S_{n}^\text{(CFT)}\right) \ .
\end{align}
As seen from \eqref{ST_one_point} the residual conformal symmetry $\SO(1,d)$ restricts the one-point function of the ambient primary operators of non-zero spin to zero.
This is also seen in \eqref{Defect_Entropy_universal} for $p=d-1$.
It is straightforward to repeat the same argument as before for BCFTs, and we are led to the results:

\bigskip
{\it
	In a DCFT with a conformal defect of dimension $d-1$ the defect entropy is given, up to UV divergence, by
	\begin{align}\label{Defect_Entropy_codim_one_universal}
		S_\mathrm{defect} = \log\, \langle\, \CD^{(d-1)}\,\rangle \ .
	\end{align}
	In a BCFT, the boundary entropy is given, up to UV divergence, by
    \begin{align}\label{Boundary_Entropy_universal}
    	S_\mathrm{bdy} = \log\, Z^\mathrm{(BCFT)} - \frac{1}{2}\,\log\, Z^\mathrm{(CFT)} \ .
    \end{align}
}

\noindent
These are the special cases of the universal formula \eqref{Defect_Entropy_universal} for the defect entropy.
In BCFT, the defect free energy is given by
\begin{align}\label{Vev_D_BCFT}
	\log\, \langle\, \CD^{(d-1)}\,\rangle \big|_\text{BCFT} = \log\, Z^\mathrm{(BCFT)} - \frac{1}{2}\,\log\, Z^\mathrm{(CFT)} \ ,
\end{align}
while in an interface CFT consisting of two theories CFT$_+$ and CFT$_-$ we define
\begin{align}
	\log\, \langle\, \CD^{(d-1)}\,\rangle \big|_\text{ICFT} = \frac{1}{2}\,\left(\log\, Z^{(\mathrm{CFT}_+)} + \log\, Z^{(\mathrm{CFT}_-)}\right)- \log\, Z^\mathrm{(CFT)} \ .
\end{align}

\subsection{UV divergence}
We turn to specify the structure of the UV divergences in the defect free energy and defect entropy.
First we note that the defect free energy should be considered as a functional of the background ambient metric and the induced metric on a defect.
In a local QFT, the UV divergent terms in the vev of a defect operator should consist of local diffeomorphism invariant functionals of the metrics on the worldvolume of the defect.
From the dimensional ground, the most general effective action for the defect vev takes the following form (see e.g. \cite{Solodukhin:2011gn,Nishioka:2018khk})
\begin{align}\label{UV_structure}
	\log\, \langle\, \CD^{(p)}\,\rangle = \int_{\CD^{(p)}}\, \d^p \hat{x}\,\sqrt{\hat{g}}\, \left[ \frac{a_p}{\epsilon^p} + \frac{a_{p-2}}{\epsilon^{p-2}}\, \hat{\CR} + \cdots \right] + (\text{UV finite non-local terms}) \ ,
\end{align}
where $\hat{\CR}$ is the Ricci scalar of the induced metric $\hat{g}$, $\epsilon\ll R$ is the UV cutoff and $a_i$ are dimensionless constants.
The $\cdots$ terms are subleading UV divergent terms built out of the Riemann curvature of the induced metric and the even power of the extrinsic curvatures.\footnote{The odd powers of the extrinsic curvatures can be added if the defect operator has the orientation specified by the normal vectors as in BCFT and ICFT.
We thank Chris Herzog for pointing this out to us.
}
For instance, the order of $1/\epsilon^{p-2i}$ divergent term roughly takes the form
\begin{align}
	\frac{a_{p-2i}}{\epsilon^{p-2i}}\,\sum_{l+m = i}\,\hat{\CR}^l\,\CK^{2m} \ ,
\end{align}
where $\hat{\CR}^l\,\CK^{2m}$ are scalar polynomials of the Riemann curvatures and the extrinsic curvatures on the defect of order $l$ and $2m$ respectively.
There are only power law divergences in odd $p$ dimensions while one can construct dimension $p$ invariants out of $\hat{\CR}$ and $\CK$ such as the Euler density and there is an additional logarithmically divergent term $\log\epsilon$.\footnote{There are also additional logarithmic divergences for odd $p$ in BCFTs \cite{Herzog:2015ioa,Fursaev:2015wpa,Fursaev:2016inw,Herzog:2017xha}.}

Applying \eqref{UV_structure} to the defect free energy on a sphere, we find the structure of the UV divergences depending on the dimensionality of the defect,
\begin{align}\label{UV_defect_FE}
	\log\, \langle\, \CD^{(p)}\,\rangle = \frac{c_{p}}{\epsilon^{p}} + \frac{c_{p-2}}{\epsilon^{p-2}} + \cdots +
    	\begin{cases}
    		(-1)^{p/2}\,B\, \log \epsilon + \cdots \ , & (p: \text{even}) \ , \\
            (-1)^{(p-1)/2}\, D \ , & (p: \text{odd}) \ .
    	\end{cases}
\end{align}
Here the sign factors in front of $B$ and $D$ are chosen so that they are non-negative.
It follows from this structure that the coefficients $c_i$ ($i=p,\,p-2,\cdots$) of the power law divergences depend on the choice of the UV cutoff and are regularization scheme dependent while the constants $B$ and $D$ are invariant under the rescaling of $\epsilon$, hence be scheme independent.
The universal constant $B$ is an analog of the type $A$ central charge of the conformal anomaly which can be read off from the sphere partition function in CFT.
It is also known as the Graham-Witten anomaly \cite{Graham:1999pm}.
Similarly $D$ is an analog of the sphere partition function that is expected to measures the degrees of freedom in CFT \cite{Giombi:2014xxa,Fei:2015oha}.

On the other hand, the UV divergent terms of the defect entropy is also inferred from the generic structure \eqref{UV_structure} with the standard argument of the replica trick \cite{Ryu:2006ef,Nishioka:2018khk}, resulting in milder divergences than the defect free energy:
\begin{align}\label{UV_defect_entropy}
	S_\text{defect} = \frac{c_{p-2}'}{\epsilon^{p-2}} + \frac{c_{p-4}'}{\epsilon^{p-4}} + \cdots +
    	\begin{cases}
    		(-1)^{p/2}\,B'\, \log \epsilon + \cdots \ , & (p: \text{even}) \ , \\
            (-1)^{(p-1)/2}\,D' \ , & (p: \text{odd}) \ ,
    	\end{cases}
\end{align}
where $B'$ and $D'$ are universal constants different from $B$ and $D$ in general.
The same UV structure was also observed in a few holographic calculations in \cite{Estes:2014hka}, where the universal constants $B'$ and $D'$ were speculated to be $C$-functions in DCFTs.

In the rest of the paper, we will study several examples in field theories and holographic models, and examine to what extent the universal constants $B,D$ and $B',D'$ are a good measure of degrees of freedom under a defect RG flow.

\section{Proposal for a $C$-theorem in DCFT}
\label{sec3}
Now we have two candidates for a $C$-function in DCFT, the universal parts of the defect entropy and the defect free energy, both of which are natural counterparts of the $C$-theorem in CFT employing the entanglement entropy across a sphere or equivalently the sphere free energy as a $C$-function  \cite{Myers:2010xs,Myers:2010tj,Casini:2011kv,Jafferis:2011zi,Klebanov:2011gs}.
The universal constants $B,D$ in \eqref{UV_defect_FE} should be regarded as analogs of the type $A$ central charge and the sphere free energy in CFT while 
the universal constants $B',D'$ in \eqref{UV_defect_entropy} differ from $B,D$ due to the relation \eqref{Defect_Entropy_universal}.
To incorporate the $b$-theorem correctly
we propose that the universal part of the defect free energy be a $C$-function in DCFT:
\begin{conjecture}\label{conjecture}
	In DCFT$_d$ with a defect of dimension $p$, the universal part of the defect free energy \eqref{Defect_Free_Energy} defined by
    \begin{align}\label{D_coefficient}
    	\tilde D \equiv \sin \left( \frac{\pi p}{2}\right)\,\log\,|\langle \, \CD^{(p)}\,\rangle|\ ,
    \end{align}
    does not increase along any defect RG flow
    \begin{align}\label{D_inequality}
    	\tilde D_\mathrm{UV} \ge \tilde D_\mathrm{IR} \ .
    \end{align}
\end{conjecture}
Notice that we take the absolute value $|\langle \, \CD^{(p)}\,\rangle|$ to define the universal part.
This is because there is a phase ambiguity in $\langle \, \CD^{(p)}\,\rangle$ such as the framing anomaly in the Chern-Simons theory which should be  removed to extract the universal part as we will encounter in section \ref{ss:Wilson}

As seen from the relations \eqref{Boundary_Entropy_universal} and  \eqref{Vev_D_BCFT}, our conjecture includes, as a special case, the statement that the universal part of the boundary entropy defined by
\begin{align}\label{D_coefficient_bdy}
    	\tilde D \equiv \sin \left( \frac{\pi (d-1)}{2}\right)\,S_\mathrm{bdy} \ ,
\end{align}
does not increase along any boundary RG flow in BCFT$_d$.

Our conjecture is the most general one in the sense that it is consistent with all the proposals stated in literatures as we will show momentarily. 

We multiply $\sin(\pi p/2)$ to the defect free energy to interpolate between $B$ for even $p$ and $D$ for odd $p$ smoothly in the dimensional regularization as in the generalized $F$-theorem \cite{Giombi:2014xxa}.
Compared with the UV divergent structure \eqref{UV_defect_FE}, $\tilde D$ is nothing but the universal part of the defect free energy for odd $p$ 
\begin{align}
	\tilde D = D  \ ,
\end{align}
while one finds a more nontrivial relation for even $p$ 
\begin{align}
	\tilde D = \frac{\pi}{2}\,B \ .
\end{align}

When $p$ is odd, our conjecture states the monotonicity of the constant universal term,
\begin{align}
D_\text{UV} \geq D_\text{IR} \ .
\end{align}
For BCFT$_2$, this is just a weak form of the $g$-theorem \cite{Affleck:1991tk,Friedan:2003yc,Casini:2016fgb}.
For BCFT$_d$ with $d\ge 3$, a similar conjecture was proposed by \cite{Nozaki:2012qd,Gaiotto:2014gha} and examined holographically in \cite{Fujita:2011fp,Estes:2014hka}.
For $d\geq 3$ and $p\le d-2$, our proposal states a new one.

When $p$ is even, our assertion derives
\begin{align}
B_\text{UV} \geq B_\text{IR} \ ,
\end{align}
which was speculated to hold in $d=3$ based on the studies of the holographic models of BCFTs and ICFTs \cite{Nozaki:2012qd,Estes:2014hka}.
For $p=2$, this is equivalent to the $b$-theorem \cite{JO1} stating the monotonicity
\begin{align}
b_{\text{UV}}\geq b_{\text{IR}}\ ,
\end{align}
of the universal coefficient $b$ of DCFT appearing in the trace of the stress-energy tensor on the defect\footnote{Our convention of the stress tensor differs from the one in \cite{JO1} up to the sign.}
\begin{align}
\langle\, t^{\mu}{}_{\mu}\, \rangle = -\frac{1}{24 \pi} \left[ b\,\hat{\mathcal{R}}+d_1\, \tilde\CK^{(\alpha)}_{ab}\tilde\CK^{(\alpha)\,ab} + d_2 \,W_{abcd}\, \hat{g}^{ac}\,\hat{g}^{bd} \right] \delta^{d-2} (x_\perp )\ ,
\end{align}
where $\tilde\CK^{(\alpha)}_{ab} \equiv \CK^{(\alpha)}_{ab} - \CK^{(\alpha)\,c}_{c}\,\hat{g}_{ab}/2$ is the traceless part of the extrinsic curvature for the normal vector $n^{(\alpha)}_a$ ($\alpha = 1, \cdots, d-2$), and $W_{abcd}$ is the pullback of the ambient Weyl tensor.
In fact, $B$ is proportional to $b$ up to a positive constant.
To fix the proportional constant one may consider a spherical defect of radius $l$ and see how the defect free energy changes under the Weyl rescaling.
Since $\tilde\CK^{(\alpha)}_{ab}$ and $W_{abcd}$ vanish on a sphere\footnote{One can show $\tilde\CK^{(\alpha)}_{ab} = 0$ by mapping the ambient sphere and the two-sphere to flat space and a two-sphere and computing the extrinsic curvatures as $\tilde\CK^{(\alpha)}_{ab}$ is conformal covariant \cite{Solodukhin:2008dh}.} the Weyl rescaling reads
\begin{align}
	l\,\frac{\d}{\d l}\log\,\langle\,\CD^{(2)}\,\rangle &= - \int \d^d x\,\sqrt{g}\,\langle\, t^{\mu}{}_{\mu}\, \rangle = \frac{b}{3} \ ,
\end{align}
which fixes the logarithmic divergent term
\begin{align}
	\log\,\langle\,\CD^{(2)}\,\rangle = \cdots + \frac{b}{3} \log \frac{l}{\epsilon} + \cdots \ ,
\end{align}
where we recover the UV cutoff $\epsilon$ to make the argument of the logarithm dimensionless.
Hence compared with \eqref{UV_defect_FE} we find
\begin{align}
	B = \frac{b}{3} \ .
\end{align}

In total, our conjecture not only unifies all the previous ones known to us, but also generates a new family of $C$-theorems in DCFTs with higher-codimensional defects.
We will provide a number of results in support of the conjecture in a variety of concrete examples in the following sections.

\subsection{Conformal perturbation theory on defect}

To examine the validity of our conjecture, we first consider the conformal perturbation theory of DCFT on a sphere, which is a straightforward extension of the works for CFT on a sphere \cite{Cardy:1988cwa,Klebanov:2011gs} and BCFT on a hemisphere \cite{Nozaki:2012qd,Gaiotto:2014gha}.
Since the calculation is exactly the same as the ambient case, just replacing the ambient dimension $d$ with the defect dimension $p$, we will only give the outline.

We locate DCFT on a sphere of a radius $R$ and perturb the theory by a defect relevant operator $\mathcal{\hat{O}}$,
\begin{align}
	I = I_\text{DCFT} + \hat \lambda_0\,\int \d^p \hat{x}\,\sqrt{\hat{g}}\,\hat \CO (\hat{x}) \ .
\end{align}
Let the conformal dimension of $\mathcal{\hat{O}}$ be $\hat{\Delta}=p-\epsilon$ and take $\epsilon$ be very small so that a nontrivial fixed point can be reliably studied within the perturbation theory.
Introducing the dimensionless renormalized coupling $\hat\lambda$ that is related to the bare coupling $\hat\lambda_0$ by 
\begin{align}
	\hat\lambda_0\,(2R)^\epsilon = \hat \lambda  + \frac{\pi^{p/2}}{\epsilon\,\Gamma (p/2)}\,\hat C\,\hat \lambda^2 + O (\hat \lambda^3) \ ,
\end{align}
the beta function is given by \cite{Cardy:1988cwa,Klebanov:2011gs}
\begin{align}
	\beta (\hat \lambda) = -\epsilon\,\hat \lambda + \frac{\pi^{p/2}}{\Gamma (p/2)}\,\hat C\, \hat \lambda^2 + O (\hat \lambda^3) \ ,
\end{align}
where  $\hat{C}$ is the coefficient appearing in the three-point function of defect local operators evaluated at the unperturbed DCFT
\begin{align}
\langle \, \mathcal{\hat{O}}(\hat{x}_1)\,\mathcal{\hat{O}}(\hat{x}_2)\, \mathcal{\hat{O}}(\hat{x}_3)\,\rangle_0 = \frac{\hat{C}}{|\hat{x}_1-\hat{x}_2|^{\hat{\Delta}}|\hat{x}_2-\hat{x}_3|^{\hat{\Delta}}|\hat{x}_3-\hat{x}_1|^{\hat{\Delta}}} \ .
\end{align}
Hence if $\hat C >0$ the theory flows to a nontrivial IR fixed point at
\begin{align}\label{Fixed_Point}
	\hat \lambda_\ast = \frac{\Gamma (p/2)}{\pi^{p/2}\,\hat C}\,\epsilon + O (\epsilon^2) \ .
\end{align}

The difference of the sphere partition function is calculated perturbatively 
\begin{align}
\delta \log\,Z (\hat\lambda) \equiv \log\, Z (\hat\lambda_0) - \log\, Z(\hat\lambda_0 =0) =\frac{\hat\lambda_0^2}{2}\,I_2 -\frac{\hat\lambda_0^3}{6}\,I_3 + O(\hat\lambda_0^4)\ ,
\end{align}
where 
\begin{align}
I_2&=\int \! \dd^p \hat{x}_1 \sqrt{\hat{g}} \int \! \dd^p \hat{x}_2 \sqrt{\hat{g}} \, \langle\,  \mathcal{\hat{O}}(\hat{x}_1)\,\mathcal{\hat{O}}(\hat{x}_2)\, \rangle_0 = \frac{\pi^{p+1/2}\,(2R)^{2\epsilon}}{2^{p-1}} \frac{\Gamma (-p/2+\epsilon)}{\Gamma \left((p+1)/2\right)\, \Gamma (\epsilon)}\ , \\
\begin{split}
I_3&=\int \! \dd^p \hat{x}_1 \sqrt{\hat{g}}\int \! \dd^p \hat{x}_2 \sqrt{\hat{g}}\int \! \dd^p \hat{x}_3 \sqrt{\hat{g}} \, \langle \, \mathcal{\hat{O}}(\hat{x}_1)\,\mathcal{\hat{O}}(\hat{x}_2) \,\mathcal{\hat{O}}(\hat{x}_3)\,\rangle_0 \\
&= \frac{8\pi^{3(p+1)/2} R^{3\epsilon}}{\Gamma (p)} \frac{\Gamma ((-p+3\epsilon)/2)}{\Gamma ((1+\epsilon)/2)^3}\, \hat{C}\ .
\end{split}
\end{align}
Written in terms of the renormalized coupling, one finds \cite{Klebanov:2011gs}
\begin{align}
	\delta \log\,Z(\hat\lambda) = \frac{2\pi^{p+1}}{\sin(\pi p/2)\,\Gamma(p+1)} \,\left[-\frac{1}{2}\,\epsilon\,\hat\lambda^2 + \frac{1}{3}\,\frac{\pi^{p/2}}{\Gamma (p/2)}\, \hat{C}\, \hat{\lambda}^3 + O(\hat{\lambda}^4) \right] \ .
\end{align}
Thus the difference between the universal part of the defect free energy at the IR fixed point \eqref{Fixed_Point} and that at the UV fixed point is
\begin{align}
	\tilde D (\hat\lambda_\ast) - \tilde D (0) = - \frac{1}{3}\,\frac{\pi\, \Gamma (p/2)^2}{\Gamma(p+1)}\,\frac{\epsilon^3}{\hat C^2} + O (\epsilon^4) \ , 
\end{align}
which is negative as consistent with our conjecture.

\subsection{Wilson loop as a defect operator}\label{ss:Wilson}
Next we test our proposal for $p=1$ using a circular Wilson loop operators
\begin{align}
 W_\mathfrak{R}[A]  = \Tr_\mathfrak{R}\,\exp \left[ \i \, \int \! \d x^\mu  A_\mu\right] \ .
\end{align}
We assume that the gauge group is $\SU(N)$ and $\mathfrak{R}$ is a representation of $\SU(N)$ for a moment.
The Wilson loop can be regarded as an action localized on the defect in the following way \cite{Gomis:2006sb,Tong:2014cha}.\footnote{See also a recent work \cite{Hoyos:2018jky} for a different formulation of a defect theory on Wilson loops.}
First we consider fermions localized on the defect and coupled to the gauge field,
\begin{align}
I_\chi=\int \! \dd t \, \chi^\dagger \left(\i\, \partial_t -A(t)\right)\chi \ ,
\end{align}
where $\chi_a$ is in the fundamental representation of $\SU(N)$.  
Then, the partition function on the defect,
\begin{align}
Z_q[A] \equiv \frac{1}{q!}\int \! \mathcal{D}\chi^\dagger \mathcal{D}\chi \, \chi_{a_1} (+\infty) \cdots \chi_{a_q}(+\infty)\, \chi^{\dagger, a_1} (-\infty) \cdots \chi^{\dagger,a_q}(-\infty) \, \mathrm{e}^{- I_\chi} \ ,
\end{align}
is equivalent to the Wilson loop up to a normalization factor
\begin{align}
\frac{Z_q[A]}{Z_q[0]} = W_\mathfrak{R}[A] \ ,
\end{align}
where the representation $\mathfrak{R}$ in the Wilson loop depends on whether $\chi$ are fermions or bosons.
When $\chi$ are fermions (bosons), $\mathfrak{R}$ is the $q^{\text{th}}$ anti-symmetric (symmetric) representation of $\SU(N)$. 

Given this description, the defect theory can flow to the trivial theory without fermions, or equivalently
\begin{align}
W_\mathfrak{R} [A] ~\to~ 1 \ ,
\end{align}
under the mass deformation
\begin{align}
I_M=-\int \! \dd t \, M\,\chi^\dagger  \chi \ ,
\end{align}
by sending $M$ to the infinity.

In what follows, we assume that any Wilson loop has a realization as a defect theory and there exists a defect RG flow whose IR fixed point is a trivial theory without loops.
Under this assumption, our conjecture amounts to the inequality
\begin{align}
	\log \,\langle\, W_\mathfrak{R}\, \rangle|_\text{UV} \ge \log \,\langle\, W_\mathfrak{R}\, \rangle|_\text{IR} = 0 \ .
\end{align}
We will provide evidences for our assertion by working out a few examples.

\subsubsection{$\U (1)$ gauge theory in 4$d$}
Our first example is the Wilson loop in a four-dimensional $\U (1)$ gauge theory
\begin{align}
	W  = \exp \left[ \i \,e \oint\,\d x^\mu A_\mu\right] \ , \qquad e \in \BR \ .
\end{align}
The defect free energy is given by
\begin{align}
		\log\,\langle\, W\, \rangle = \frac{e^2}{4} \ ,
\end{align}
which is seen to be positive while the defect entropy vanishes \cite{Lewkowycz:2013laa}
\begin{align}
	S_\text{defect} = 0 \ .
\end{align}
It is expected that the Wilson loop becomes trivial under a defect RG flow,
\begin{align}
\log\, \langle\, W\, \rangle \to 0 \ ,
\end{align}
so this is consistent with our conjecture.
On the other hand, the defect entropy vanishes at both the UV and IR fixed points.
Hence, the defect entropy does not appear to capture degrees of freedom on the defect.

\subsubsection{Free scalar field in 4$d$}
The next example is a scalar Wilson loop in four dimensions \cite{Kapustin:2005py}
\begin{align}
	W  = \exp \left[ \lambda\oint\,\d t\, \phi \left(x^\mu(t)\right)\right] \ ,\qquad \lambda \in \BC \ .
\end{align}
The defect free energy is computed by evaluating the Gaussian integral, and shown to vanish
\begin{align}
	\log\,\langle\, W\, \rangle = 0 \ .
\end{align}
Reassuringly this result does not contradict with our assertion.
On the other hand, the defect entropy is given by \cite{Lewkowycz:2013laa}
\begin{align}
S_{\text{defect}}=-\frac{\lambda^2}{12} \ ,
\end{align}
which can be negative for real $\lambda$ at the UV fixed point
while it is supposed to be zero at the IR fixed point.
Thus this is a counterexample for the defect entropy being a $C$-function.

\subsubsection{Chern-Simons theory}
As a more nontrivial example, let us consider Wilson loops in the Chern-Simons theory in three-dimensions\footnote{Note that our normalization for Wilson loops are different from the one in \cite{Kapustin:2009kz} where the operators are divided by the dimension of the representation.}
\begin{align}
	W_\mathfrak{R} = \Tr_\mathfrak{R}\, \CP\,\exp \left[ \i\,\oint \d x^\mu A_\mu \right] \ .
\end{align}
For $\SU (2)$ with level $k$, the Wilson loop in the representation $\mathfrak{R}_j$ is labeled by the dimension $j=1, \cdots, k+1$, whose vev on $\BS^3$ is \cite{Witten:1988hf,Beasley:2009mb}
\begin{align}
        	\langle\, W_{\mathfrak{R}_j}\, \rangle = \frac{\sin\left( \pi\,j/ (k+2)\right)}{\sin\left( \pi / (k+2)\right)} \ ,
\end{align}
which is greater than or equal to one.

More generally the vev of a Wilson loop in an arbitrary representation $\mathfrak{R}_j$ on $\BS^3$ is given by \cite{Witten:1988hf}
\begin{align}
	\langle\, W_{\mathfrak{R}_j}\, \rangle = \frac{S_{0,j}}{S_{0,0}} \equiv d_j \ ,
\end{align}
where $S_{i, j}$ is the matrix element of the modular group $S$-matrix.
The vev or $d_j$ is called the quantum dimension of $\mathfrak{R}_j$, which is known to be greater than or equal to one \cite{Dijkgraaf:1988tf} (and see also Appendix C in \cite{Shi:2018bfb}),
\begin{align}
	d_j \ge 1 \ .
\end{align}
This is consistent with our conjecture.
Note that the defect entropy is also given by
\begin{align}
	S_\text{defect} = \log\,\langle\, W_{\mathfrak{R}_j}\, \rangle = \log\,d_j \ ,
\end{align}
as the stress tensor vanishes in Chern-Simons theory.\footnote{This result was previously obtained by \cite{Dong:2008ft,Balasubramanian:2016sro,Wong:2017pdm}.}

\subsubsection{$1/2$-BPS Wilson loop in 4$d$ $\CN =4$ SYM}
\begin{figure}[htbp]
	\centering
	\includegraphics[width=7.5cm]{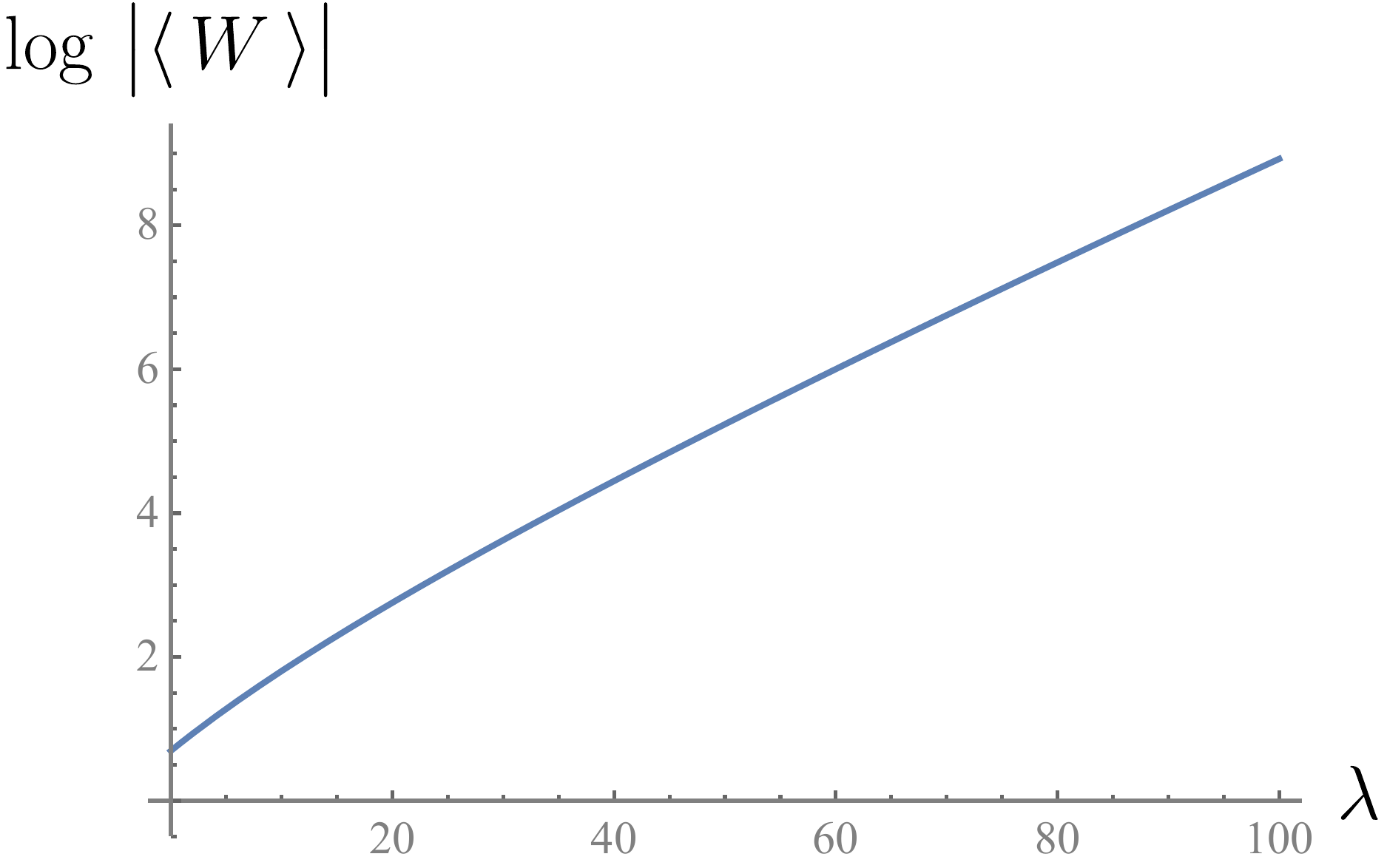}
    ~~
	\includegraphics[width=7.5cm]{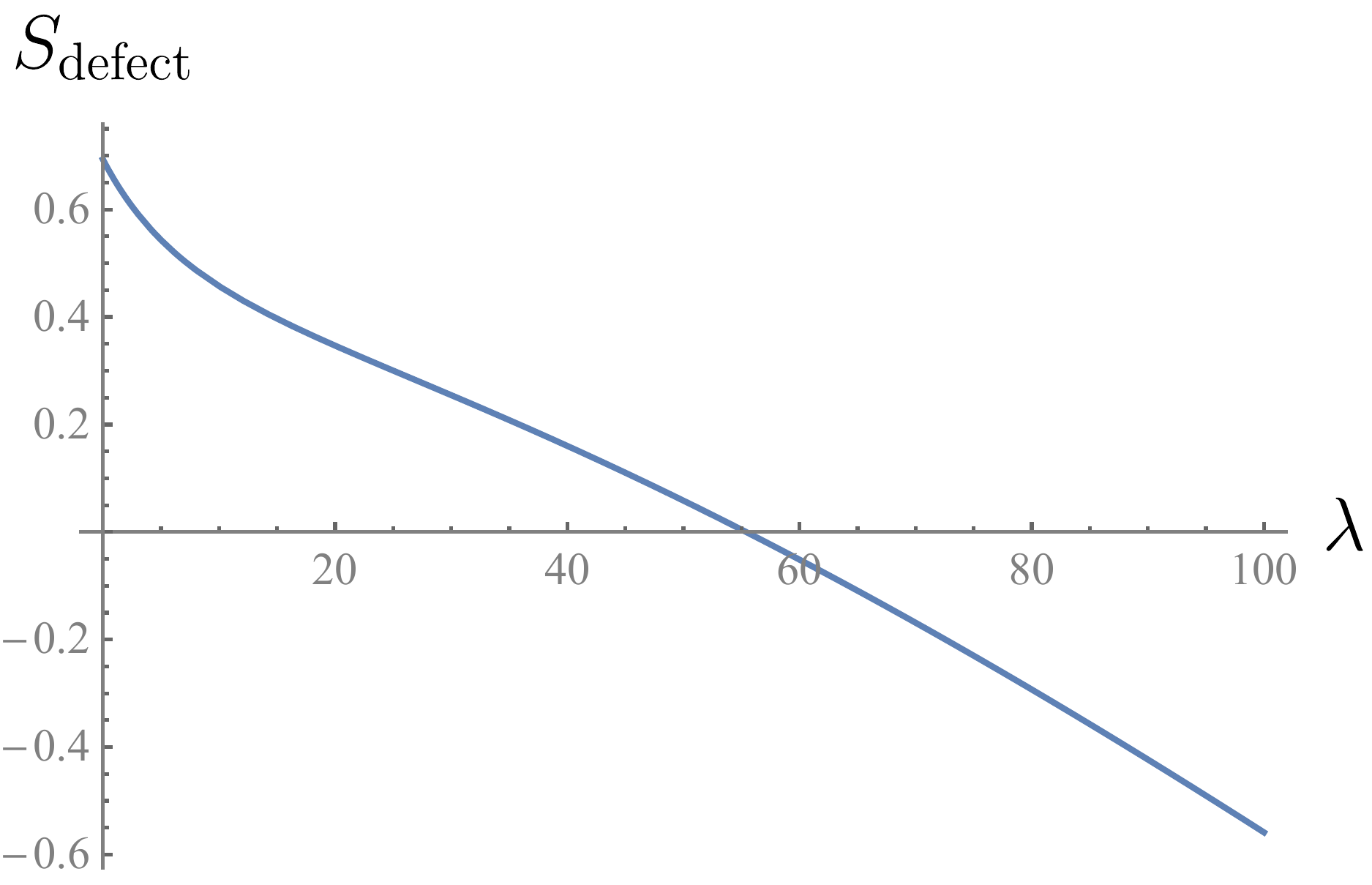}
    \caption{The defect free energy (Left) and the defect entropy (Right) of the 1/2-BPS Wilson loop in the 4$d$ $\CN=4$ SYM.
    The $N=2$ cases are shown.
    The defect free energy is positive for any $\lambda$ while the defect entropy can be negative.}
    \label{fig:WilsonLoop}
\end{figure}

There are the $1/2$-BPS Wilson loops in the four-dimensional $\CN =4$ super Yang-Mills theory with gauge group $\U (N)$
\begin{align}\label{1/2BPSWL}
	W_\mathfrak{R} = \Tr_\mathfrak{R}\, \CP\, \exp \left[ \oint \d t \,( \i\, A_\mu \,\dot x^\mu + \phi_I\,\dot y^I )\right] \ .
\end{align}
For the fundamental representation, the exact result of the defect free energy  is known \cite{Drukker:2000rr}
\begin{align}
	\log\,\langle\, W\, \rangle = \frac{\lambda}{8N} + \log L^1_{N-1} \left( -\frac{\lambda}{4N}\right)\ ,
\end{align}
where $L^m_n (x)$ is the associated Laguerre polynomial.
In the small $\lambda$ region we find the expansion
\begin{align}
	\log\,\langle\, W\, \rangle = \log N + \frac{\lambda}{8} - \frac{1}{384}\left( 1- \frac{1}{N^2}\right)\lambda^2 + O(\lambda^3) \ ,
\end{align}
which is seen to be positive for any $N$ and small $\lambda$.
One can indeed check numerically it is always positive for any $N$ and $\lambda$ (see the left panel in figure \ref{fig:WilsonLoop}).

On the other hand, the defect entropy can be calculated from the defect free energy through the relation \cite{Lewkowycz:2013laa}
\begin{align}\label{WL_DE}
	S_\text{defect} = \left( 1 - \frac{4}{3}\lambda\, \partial_\lambda \right)\, \log\,\langle\, W\, \rangle \ .
\end{align}
Then we find that the entropy is not necessarily positive in the small $\lambda$ limit (see also the right panel in figure \ref{fig:WilsonLoop})
\begin{align}
	S_\text{defect} = \log N - \frac{\lambda}{24} + \frac{5}{1152}\left( 1- \frac{1}{N^2}\right)\lambda^2 + O(\lambda^3) \ .
\end{align}
This example also serves as a supporting evidence for our conjecture and a nontrivial counterexample for the defect entropy being a $C$-function.

\paragraph{RG flow from non-supersymmetric Wilson loop to $1/2$-BPS Wilson loop}
The $1/2$-BPS Wilson loop \eqref{1/2BPSWL} is a special case of the generalized Wilson loop \cite{Polchinski:2011im}
\begin{align}
	W^{(\zeta)} = \Tr\, \CP\, \exp \left[ \oint \d t \,( \i\, A_\mu \,\dot x^\mu + \zeta\,\phi_I\,\dot y^I )\right] \ ,
\end{align}
with a constant parameter $\zeta$.
It is supersymmetric only when $\zeta = 1$ and reduces to the standard Wilson loop when $\zeta =0$.
The $\zeta$-dependent term is a weakly relevant perturbation on the loop, which triggers an RG flow from the standard Wilson loop at the UV fixed point to the $1/2$-BPS Wilson loop at the IR fixed point.
This flow has been studied extensively by \cite{Beccaria:2017rbe} at both week and strong coupling $\lambda$ in the large $N$ limit.

At weak coupling, one finds \cite{Beccaria:2017rbe}
\begin{align}
	\log\,\langle\, W^{(\zeta)}\, \rangle = \log N + \frac{\lambda}{8} - \frac{1}{384}\left( 1- \frac{3(1-\zeta^2)^2}{\pi^2}\right)\lambda^2 + O(\lambda^3) \ ,
\end{align}
which is consistent with our conjecture,\footnote{We thank Simone Giombi and Arkady Tseytlin for informing us of their relevant work.}
\begin{align}
	\log\,\langle\, W^{(0)}\, \rangle > \log\,\langle\, W^{(1)}\, \rangle \ .
\end{align}
On the other hand, the defect entropy calculated by \eqref{WL_DE}\footnote{The formula \eqref{WL_DE} works only at the fixed point, hence this is not the defect entropy of the generalized Wilson loop unless $\zeta = 0, 1$.}
\begin{align}
	S_\text{defect}^{(\zeta)} = \log N - \frac{\lambda}{24} + \frac{5}{1152}\left( 1- \frac{3(1-\zeta^2)^2}{\pi^2}\right)\lambda^2 + O(\lambda^3) \ ,
\end{align}
increases along the flow,
\begin{align}
	S_\text{defect}^{(0)} < S_\text{defect}^{(1)}  \ .
\end{align}

Our conjecture is also consistent with the strong coupling result in \cite{Beccaria:2017rbe}.

\paragraph{RG flows interpolating between various representations}
Two RG flows interpolating between the $1/2$-BPS Wilson loops in different representations are considered in \cite{Kumar:2016jxy,Kumar:2017vjv}\footnote{We thank Prem Kumar for informing us of the relevant works and for helpful correspondences.}:
\begin{itemize}
	\item an RG flow from the $k$ fundamental representation to the anti-symmetric representation,
	\item an RG flow from the symmetric representation to the $k$ fundamental representation.
\end{itemize}
These flows are constructed holographically by D-brane probes, which allow us to calculate the defect free energies and the defect entropies in the $N\to \infty$ limit with $k/N$ fixed at strong coupling.
It is shown that the defect free energies decrease in both cases while the defect entropy increases in the latter \cite{Kumar:2016jxy,Kumar:2017vjv}.

\subsubsection{1/6-BPS Wilson loop in ABJM}
There is the 1/6-BPS Wilson loop in the fundamental representation of the ABJM theory in three dimensions with gauge groups $\U (N)_k\times \U(N)_{-k}$,
\begin{align}
			W= \Tr\, \CP\, \exp \left[ \oint \d t \,( \i\, A_\mu \,\dot x^\mu + \frac{2\pi}{k} M^I_J\,C_I\,C^J\,|\dot x| )\right] \ ,
\end{align}
where $C^I (I=1,2,3,4)$ are the scalar fields in the bi-fundamental chiral multiplets and $M^I_J$ is a constant matrix whose diagonalized form is $\text{diag}(1,1,-1,-1)$.
The supersymmetric localization allows us to compute the vev of the $m$ multiply-winding Wilson loop $W^{(m)}$ by the matrix model \cite{Kapustin:2009kz}
\begin{align}
\begin{aligned}
	\langle \, W^{(m)} \, \rangle &= \frac{1}{Z}\,\frac{1}{(N!)^2}\,\int \prod_{i=1}^N\,\frac{\d \mu_i\,\d \nu_i}{(2\pi)^2}\, \mathrm{e}^{\i\,k(\mu_i^2 - \nu_i^2)/4\pi}\,
    \\
 & \qquad  \cdot  \frac{\prod_{i<j} \left[4\sinh\left( (\mu_i - \mu_j)/2\right)\,\sinh\left((\nu_i - \nu_j)/2\right) \right]^2}{\prod_{i,j} \left[ 2\cosh \left( (\mu_i - \nu_j)/2\right)\right]^2}\,\sum_i\, \mathrm{e}^{m\mu_i} \ ,
\end{aligned}
\end{align}
where $Z$ is the partition function 
\begin{align}
	Z = \frac{1}{(N!)^2}\,\int \prod_{i=1}^N\,\frac{\d \mu_i\,\d \nu_i}{(2\pi)^2}\, \mathrm{e}^{\i\,k(\mu_i^2 - \nu_i^2)/4\pi}\,\frac{\prod_{i<j} \left[4\sinh\left( (\mu_i - \mu_j)/2\right)\,\sinh\left((\nu_i - \nu_j)/2\right) \right]^2}{\prod_{i,j} \left[ 2\cosh \left( (\mu_i - \nu_j)/2\right)\right]^2} \ .
\end{align}
Performing the integral exactly is quite difficult in general, but it is straightforward for $N=1$,
\begin{align}\label{ABJM_multiple}
	\langle \, W^{(m)} \, \rangle = \cos^{-2} \left( \frac{\pi m}{k}\right) \ , \qquad\qquad (N=1) \ .
\end{align}
This is seen to be greater than or equal to one for $m=1$ and any $k$, hence consistent with our conjecture.

The defect entropy can be read off from the vev of the winding Wilson loop by the formula
\begin{align}\label{DefEnt_3d}
	S_\text{defect} = \lim_{m\to 1}\,\left( 1 - \frac{1}{2}\,m \partial_m \right) \, \log|\langle\, W^{(m)}\,\rangle| \ ,
\end{align}
which is derived in \cite{Lewkowycz:2013laa} using the supersymmetric R\'enyi entropy \cite{Nishioka:2013haa}.
Substituting \eqref{ABJM_multiple} into \eqref{DefEnt_3d} we find
\begin{align}
	S_\text{defect} = - \log \cos^2 \left(\frac{\pi}{k}\right) - \frac{\pi}{k} \tan \left( \frac{\pi}{k} \right) \ ,
\end{align}
which is always negative for positive integer $k$.

In the large $N$ limit, the matrix model reduces to the integral  \cite{Marino:2009jd}
\begin{align}
	\langle\, W^{(m)}\,\rangle = \frac{N}{2\pi^2\,\i\, \lambda}\,\int_{-a}^a\, \d x\, \mathrm{e}^{mx}\,\arctan\,\sqrt{\frac{\alpha - 2\cosh x}{\beta + 2\cosh x}} \ ,
\end{align}
where 
\begin{align}
	\mathrm{e}^a = \frac{2 + \i\, \kappa + \sqrt{\kappa (4\i - \kappa)}}{2} \ , \qquad \alpha = 2 + \i\,\kappa\ , \qquad \beta = 2 - \i\,\kappa\ ,
\end{align}
and
\begin{align}
	\lambda = \frac{N}{k} = \frac{\kappa}{8\pi}\, {}_3F_2 \left( \frac{1}{2}, \frac{1}{2}, \frac{1}{2}; 1, \frac{3}{2}; -\frac{\kappa^2}{16}\right) \ .
\end{align}
In the small $\lambda$ limit, we find the vev of the fundamental Wilson loop \cite{Marino:2009jd}\footnote{The explicit expression for the Wilson loop valid for any $\lambda$ in the large $N$ limit was obtained in \cite{Bianchi:2018scb}.}
\begin{align}
	\log\,|\langle\, W\,\rangle| = \log N + \frac{5\pi^2\lambda^2}{6} + O(\lambda^4) \ ,
\end{align}
and the defect entropy \cite{Lewkowycz:2013laa}
\begin{align}
	S_\text{defect} = \log N - \frac{\pi^2\lambda^2}{6} + O(\lambda^4) \ ,
\end{align}
both of which are dominated by $\log N$, hence positive.
They are also increasing functions for $\lambda$ large enough.

\subsubsection{$\U (N)$ $\CN =4$ SYM with $N_f$ hypermultiplets in $3d$}
Wilson loops in three-dimensional $\CN =2$ supersymmetric theories are defined by
\begin{align}
     W_\mathfrak{R} \equiv \Tr_\mathfrak{R}\, \CP\, \exp \left[ \oint \d t \left( \i\, A_\mu\, \dot x^\mu + \sigma\, |\dot x|\right) \right] \ ,
\end{align}
where $\sigma$ is the adjoint scalar field in the vector multiplet.

As a concrete example, consider the Wilson loop in $\U (N)$ $\CN =4$ supersymmetric gauge theory with $N_f$ hypermultiplets.
The partition function in this theory is given by
\begin{align}
	Z = \frac{1}{N!}\int \prod_{i=1}^N\frac{\d \mu_i}{2\pi}\, \frac{\prod_{i< j}\,4\sinh^2 \left( (\mu_i - \mu_j)/2\right)}{\prod_i \left[ 2\cosh \left( \mu_i/2\right) \right]^{N_f}} \ ,
\end{align}
and the Wilson loop in the representation labeled by the Young diagram of the partition $\lambda$ is
\begin{align}
	\langle\, W_\lambda\,\rangle = \frac{1}{Z}\,\frac{1}{N!}\int \prod_{i=1}^N\frac{\d \mu_i}{2\pi}\, \frac{s_\lambda (\mathrm{e}^{\mu_1}, \cdots, \mathrm{e}^{\mu_N})}{\prod_i \left[ 2\cosh \left( \mu_i/2\right) \right]^{N_f}}\,\prod_{i< j}\,4\sinh^2 \left( (\mu_i - \mu_j)/2\right) \ ,
\end{align}
where $s_\lambda$ is the Schur polynomial.
This integral can be performed exactly, resulting in the simple formula \cite{Tierz:2018fsn},
\begin{align}\label{WilsonLoop_Schur_Formula}
	\langle\, W_\lambda\,\rangle = \frac{s_\lambda (1_{N_f/2})\, s_\lambda (1_N)}{s_{\lambda'} (1_{N_f/2 - N})} \ ,
\end{align}
where $\lambda'$ is the conjugate representation of $\lambda$ and 
\begin{align}
	s_\lambda (1_n) \equiv s_\lambda (1, 1, \cdots, 1) = \prod_{1\le i < j \le n}\,\frac{\lambda_i - \lambda_j + j - i}{j-i} \ .
\end{align}
Then the Wilson loop in the fundamental representation becomes
\begin{align}
	\langle\, W_{(1)}\,\rangle = \frac{N_f}{ N_f/N -2} \ ,
\end{align}
which is greater than one when $N_f > 2N$.
This regime corresponds to ``good" or ``ugly" theories while $N\le N_f < 2N$ corresponds to ``bad" theories with unitarity violating monopole operators.
In the latter parameter region, the theory is proposed to be dual to the ``good" theory of $\U(N_f - N)$ gauge group with $N_f$ hypermultiplet and $2N-N_f$ additional free (twisted) hypermultiplets \cite{Yaakov:2013fza}.

For a multiply-winding Wilson loop with winding number $m$, we replace $\mathrm{e}^{\mu_i}$ with $\mathrm{e}^{m \mu_i}$ in the argument of the Schur polynomial
\begin{align}
	\langle\, W_\lambda^{(m)}\,\rangle = \frac{1}{Z}\,\frac{1}{N!}\int \prod_{i=1}^N\frac{\d \mu_i}{2\pi}\, \frac{s_\lambda (\mathrm{e}^{m\mu_1}, \cdots, \mathrm{e}^{m \mu_N})}{\prod_i \left[ 2\cosh \left( \mu_i/2\right) \right]^{N_f}}\,\prod_{i< j}\,4\sinh^2 \left( (\mu_i - \mu_j)/2\right) \ .
\end{align}
This expression can be expanded by a linear combination of singly winding Wilson loops.
For instance, the Wilson loop with winding number $m$ in the fundamental representation 
\begin{align}
	\langle\, W_{(1)}^{(m)}\,\rangle = \sum_{l=0}^{m}\, (-1)^l\, \langle\, W_{\left( m-l,\, 1^l\right)}\,\rangle \ ,
\end{align}
which follows from the identity
\begin{align}
	s_{(1)}(x_1^m, \cdots, x_N^m) = \sum_{l=0}^{m}\, (-1)^l\, s_{\left( m-l, \,1^l\right)}(x_1,\cdots, x_N) \ .
\end{align}
With the aid of the formula \eqref{WilsonLoop_Schur_Formula} we find
\begin{align}
	\langle\, W_{\left( m-l,\, 1^l\right)}\,\rangle =
    	\frac{\Gamma(N_f/2 + m-l)\,
        	\Gamma(N+m-l)\,
            \Gamma(N_f/2 -N-m+l+1)}{m\,
            \Gamma(m-l)\,
            \Gamma(l+1)\,
            \Gamma(N_f/2-l)\,
            \Gamma(N-l)\,
            \Gamma(N_f/2-N+l+1)} \ . 
\end{align}
It follows that the vev of the winding Wilson loop is given exactly for $N=1$ by
\begin{align}
	\langle\, W_{(1)}^{(m)}\,\rangle = \frac{
    \Gamma\left( N_f/2-m\right) 
    \Gamma\left(m+N_f/2\right)}{
    \Gamma\left(N_f/2\right)^2} \ ,
\end{align}
and for $N=2$ by
\begin{align}
	\langle\, W_{(1)}^{(m)}\,\rangle = \frac{\left(N_f+2m^2-2\right) 
    \Gamma\left(N_f/2 -m-1\right) 
    \Gamma\left(N_f/2 +m-1\right)}{
    \Gamma\left(N_f/2-1\right) 
    \Gamma\left(N_f/2\right)} \ .
\end{align}
Using the expression \eqref{DefEnt_3d} for the defect entropy we obtain for $N=1$
\begin{align}
	S_\text{defect} = \log \left( \frac{N_f}{N_f-2}\right) - \frac{2 (N_f -1) }{N_f (N_f - 2)} \ ,  \qquad\qquad (N=1) \ ,
\end{align}
which is negative for $N_f> 2N = 2$, while for $N=2$ we find
\begin{align}
	S_\text{defect} =  \log \left( \frac{ 2N_f }{ N_f - 4 }\right) - \frac{2(2N_f^2 - 9 N_f + 8)}{N_f (N_f - 2)(N_f -4)}\ ,  \qquad\qquad (N=2) \ ,
\end{align}
which is positive for $N_f> 2N = 4$.
We thus conclude that the defect entropy does not necessarily decrease under the defect RG flow to the trivial fixed point in this theory.

\section{Holographic models of DCFTs}
\label{sec4}
In this section we consider a class of holographic models of DCFTs where a defect RG flow is triggered geometrically by a deformation of the spacetime.
After realizing the CHM map as a coordinate transformation in the bulk spacetime following \cite{Jensen:2013lxa} we calculate the defect entropy as the black hole entropy of the mapped spacetime.
Along the way we point out the difference between the defect entropy and the defect free energy that is holographically given by minus the on-shell action.
We then perform the holographic calculations of the defect free energy and the defect entropy in these models.
Furthermore, we establish the holographic $C$-theorem in DCFT by imposing the null energy condition on the bulk theories, which proves our conjecture in the holographic systems we study.

\subsection{CHM map and defect entropy in holography}
A general metric of an asymptotically AdS space preserving the $\SO(2,p) \times \SO(d-p)$ symmetry of DCFT takes the following form, 
\begin{align}
\dd s^2 = L^2\left[ \dd \rho^2 + A(\rho)^2\, \dd s_{\mathrm{AdS}_{p+1}}^2+B(\rho)^2\, \dd s_{\mathbb{S}^{d-p-1}}^2 \right] \ .
\label{deform}
\end{align}
For $p<d-1$ the range of $\rho$ is $0 \leq \rho < \infty$.
$A(\rho)$ and $B(\rho)$ are arbitrary positive definite functions that have the asymptotic forms near the boundary ($\rho\to \infty$)
\begin{align}
A(\rho) , \, B(\rho) ~ \to~ \frac{\exp (\rho-c_p)}{2} \ .
\end{align}
For $d=p-1$, $\rho\in (-\infty, \infty)$ and the conformal boundary sits at $\rho\to \pm \infty$.

The boundary spacetime of \eqref{deform} reached by the $\rho \to \infty$ limit is AdS$_{p+1} \times \BS^{d-p-1}$, which is conformally equivalent to $\BR \times \BH^{d-1}$ by the CHM map as expected, but one can realize such a conformal transformation more directly in the bulk by choosing the AdS topological black hole coordinates for the AdS$_{p+1}$ subspace
\begin{align}
\dd s_{\mathrm{AdS}_{p+1}}^2 = -f(V)\,\dd \tau ^2 + \frac{\dd V^2}{f(V)} +V^2\, \dd s_{\mathbb{H}^{p-1}}^2 \ ,
\end{align}
with
\begin{align}
	f(V) = V^2-1\ .
\end{align}
The resulting metric is an asymptotically AdS black hole solution with the horizon located at $V=1$ and the Hawking temperature $T_0=1/2\pi$,
\begin{align}\label{AdS_topBH}
\dd s^2 = L^2\,A(\rho)^2 \left[ -f(V)\,\dd \tau ^2 + \frac{\dd V^2}{f(V)} +V^2\, \dd s_{\mathbb{H}^{p-1}}^2  \right]+L^2\left(\dd \rho^2 +B(\rho)^2\,\dd s_{\mathbb{S}^{d-p-1}}^2\right) \ ,
\end{align}
whose asymptotic boundary at $\rho \to \infty$ becomes $\BR \times \BH^{d-1}$ of the form \eqref{Hyperbolic} up to a conformal factor
\begin{align}
	\begin{aligned}
	\dd s^2 ~ &\to \frac{L^2}{4}\,\mathrm{e}^{2(\rho-c_p)}\,f(V)\,\left[ - \dd \tau^2 + \frac{\dd V^2}{f(V)^2}+ \frac{V^2}{f(V)}\, \dd s_{\mathbb{H}^{p-1}}^2 + \frac{1}{f(V)}\,\dd s_{\mathbb{S}^{d-p-1}}^2 \right] \\
    	&= \frac{L^2}{4}\, \mathrm{e}^{2(\rho-c_p)}\,f(V)\,\left[ - \dd \tau^2 + \dd x^2 + \cosh^2 x \, \dd s_{\mathbb{H}^{p-1}}^2 + \sinh^2x \,\dd s_{\mathbb{S}^{d-p-1}}^2 \right] \ ,
    \end{aligned}
\end{align}
where we introduced the new coordinate $x$ by $V = \coth x$.

Now we want to evaluate the entanglement entropy of a spherical entangling region considered in section \ref{sec2} holographically.
There are two ways to calculate the entanglement entropy that yield the same answer: (1) use the Ryu-Takayanagi formula of the holographic entanglement entropy \cite{Ryu:2006bv}, (2) use the CHM map and equate the entanglement entropy with the thermal entropy.

Let us start with describing the first method.
In the topological black hole coordinates \eqref{AdS_topBH} 
the Ryu-Takayanagi minimal surface coincides with the black hole horizon \cite{Jensen:2013lxa}, so the entanglement entropy is given by the black hole entropy
\begin{align}\label{Hol_EE_DCFT}
S^{\text{(DCFT)}}=\frac{A_\text{H}}{4G_{\text{N}}} \ ,
\end{align}
where $A_\text{H}$ is the area of the horizon
\begin{align}\label{area}
\begin{aligned}
A_{\mathrm{H}} &=L^{d-1}\,\text{Vol} (\mathbb{S}^{d-p-1})\,\text{Vol} (\BH^{p-1})\, \int_0^\infty \! \dd \rho \, A(\rho)^{p-1}\,B(\rho)^{d-p-1} \\
	&= L^{d-1}\,\frac{2\pi^{d/2}}{\sin \left(\pi p/2\right)\Gamma \left( p/2 \right)\,\Gamma \left( (d-p)/2\right)}\, \int_0^\infty \! \dd \rho \, A(\rho)^{p-1}\,B(\rho)^{d-p-1} \ ,
\end{aligned}
\end{align}
and we used the sphere volume and the regularized volume of the hyperbolic space
\begin{align}\label{Hyp_Volume}
	\text{Vol}(\BS^{d-p-1}) = \frac{2\pi^{(d-p)/2}}{\Gamma \left( (d-p)/2\right)}\ , \qquad \text{Vol}(\mathbb{H}^{p-1}) = \frac{\pi^{p/2}}{\sin \left(\pi p/2\right)\Gamma \left( p/2 \right)} \ .
\end{align}
The defect entropy can be easily obtained in this setup.
Subtracting the holographic entanglement entropy without defect given by \eqref{Hol_EE_DCFT} with $A(\rho)=\cosh \rho$ and $B(\rho)=\sinh \rho$, one finds the holographic defect entropy
\begin{align}
\begin{aligned}
	S_{\text{defect}} &= \frac{L^{d-1}}{4G_{\text{N}}} \,\frac{2\pi^{d/2}}{\sin \left(\pi p/2\right)\Gamma \left( p/2 \right)\,\Gamma \left( (d-p)/2\right)}\\
    	&\qquad \cdot  \int_0^\infty \! \dd \rho \, \left(A(\rho)^{p-1}B(\rho)^{d-p-1}-\cosh^{p-1} \rho \,\sinh ^{d-p-1}\rho \right) \ .
\end{aligned}
\end{align}

Next we want to calculate the thermal entropy for DCFT on $\BH^{d-1}$ at finite temperature $T$ holographically that reduces to the entanglement entropy when $T = T_0$.
To this end, we replace the function $f(V)$ appeared in the AdS topological black hole metric \eqref{AdS_topBH} with 
\begin{align}\label{lapse}
	f(V) = V^2 - 1 - \frac{V_\text{H}^{p-2}}{V^{p-2}}\,(V_\text{H}^2 - 1) \ .
\end{align}
The resulting geometry is an asymptotically AdS black hole whose boundary is $\BR \times \BH^{d-1}$ that the dual DCFT lives on at temperature 
\begin{align}
	T = \frac{1}{4\pi}\,\left( p\,V_\text{H} - \frac{p-2}{V_\text{H}}\right) \ .
\end{align}
The thermal entropy is given by the black hole entropy
\begin{align}
	S_\text{thermal}(T) = V_\text{H}^{p-1}\,\frac{A_\text{H}}{4G_\text{N}} \ ,
\end{align}
which obviously reproduces \eqref{Hol_EE_DCFT} in $T\to T_0$ ($V_\text{H} \to 1$).
One can also calculate the R\'enyi entropy from the thermal entropy \cite{Hung:2011nu}
\begin{align}
	\begin{aligned}
	S_n^\text{(DCFT)} &= \frac{n}{n-1}\,\frac{1}{T_0}\,\int_{T_0/n}^{T_0}\,\dd T\, S_\text{thermal}(T)  \\
    	&=  \frac{n}{n-1}\,(2-v^{p-2} - v^p)\,\frac{A_\text{H}}{8G_\text{N}} \ ,
    \end{aligned}
\end{align}
where $v \equiv \left( 1+\sqrt{1 + p\,n^2(p-2)} \right)/p\,n$.

We can account for the difference between the defect entropy and the on-shell action in a similar manner to the field theory case. 
Suppose the holographic models of DCFT and CFT are described by the actions $I_{\text{DCFT}}[G_{MN}]$ and $I_{\text{CFT}}[G_{MN}^{(0)}]$ respectively.
Here $G_{MN}$ is the backreacted metric of the form \eqref{AdS_topBH} with \eqref{lapse} and $G_{MN}^{(0)}$ is the one with $A(\rho) =\cosh \rho$ and $ B(\rho) = \sinh \rho$.
The thermodynamic relation allows us to compute the defect contribution to the thermal entropy
\begin{align}\label{DE_brane}
	S_\text{defect} &= \lim_{T \to T_0}\left[ - \frac{\partial}{\partial T} \left( T\, \Delta I \right)\right] \notag \\
    	&= \lim_{T \to T_0}\left[ - \Delta I -  T\,\frac{\partial}{\partial T}\, \Delta I \right]\ .
\end{align}
The first term in the right hand side is the difference of the on-shell actions 
\begin{align}
	\Delta I \equiv I_{\text{DCFT}}[G_{MN}] - I_{\text{CFT}}[G_{MN}^{(0)}] \ .
\end{align}
Compared with the CFT result on the defect entropy \eqref{Defect_Entropy_universal},
we find that the first term in \eqref{DE_brane} should be identified with the defect free energy while the second term corresponds to the integrated one-point function $\int \langle\,(T_{\text{DCFT}})^\tau_{~\tau}\,\rangle$ in the dual DCFT through the GKP-W relation.
We note that there are some cases where $I_{\text{DCFT}} = I_{\text{CFT}}$. 
For example, a holographic dual of a Janus interface CFT is described by  the type IIB supergravity where the Janus interface is implemented by a nontrivial profile of the dilaton field that backreacts to the metric.
Hence $\Delta I$ is the difference between the same actions evaluated on the nontrivial and trivial profiles \cite{Bak:2003jk}.

\subsection{Domain wall defect RG flow}
The next example we consider is a holographic model of a defect RG flow interpolating between two fixed points described by the metric \eqref{deform} with the defining functions $A_\text{UV}(\rho)$, $B_\text{UV}(\rho)$ at the UV fixed point obeying the boundary conditions
\begin{align}
	A_\text{UV}(\rho),\, B_\text{UV}(\rho) ~\to~ \frac{\exp (\rho-c_\text{UV})}{2} \ ,
\end{align}
and $A_\text{IR}(\rho)$, $B_\text{IR}(\rho)$ at the IR fixed point obeying 
\begin{align}
	A_\text{IR}(\rho),\, B_\text{IR}(\rho) ~\to~ \frac{\exp (\rho-c_\text{IR})}{2} \ .
\end{align}
This is the most general situation, but we restrict our attention to the RG flow with the IR fixed point characterized by 
\begin{align}
	A_\text{IR}(\rho) = \ell\, A_\text{UV}(\rho) \ , \qquad B_\text{IR}(\rho) = \ell\, B_\text{UV}(\rho) \ ,
\end{align}
for a positive dimensionless constant $\ell$.
In this case the interpolating metric between the two fixed points must respect the Poincar\'e symmetry on and the rotational symmetry around the defect, resulting in the domain wall type ansatz
\begin{align}
\dd s ^2 =L^2 \left[ \dd \rho^2 + \frac{A_\text{UV}(\rho)^2}{f(w)}\, \dd s_{\mathrm{AdS}_{p+1}}^2 + \frac{B_\text{UV}(\rho)^2}{f(w)}\,\dd s_{\mathbb{S}_{d-p-1}}^2 \right] \ ,
\end{align}
where $w$ is the radial direction in Poincar\'{e} coordinate of the sliced AdS space,
\begin{align}
 \dd s_{\mathrm{AdS}_{p+1}}^2 = \frac{\dd w^2 -\dd t^2 + \sum_{a=1}^{p-1}\dd \hat{x}_a^2 }{w^2} \ .
\end{align}
We regard $w$ as the holographic renormalization scale ranging from the UV at $w=0$ to the IR at $w=\infty$, and impose the boundary condition
\begin{align}
	f(w) ~\to ~ 1 \ , \qquad w\, \to\, 0  \ ,
\end{align}
at the UV fixed point and
\begin{align}
	f(w) ~\to ~ \ell^{-2}  \ , \qquad w\, \to\, \infty  \ ,
\end{align}
at the IR fixed point.

In order to make the ansatz physically sensible in the Einstein gravity coupled to matters we impose the null energy condition for the matters
\begin{align}
	T_{MN} \zeta^M \zeta^N \ge 0 \ ,
\end{align}
for any null vector $\zeta^M$.
Choosing $\zeta^M$ to be $\zeta^w=1$, $\zeta^t=1$ and $\xi^{M\neq w, t} = 0$ and using the Einstein equation
\begin{align}
8\pi G_\text{N}\, T_{MN} =\mathcal{R}_{MN}-\frac{1}{2}
\,G_{MN}  \,\mathcal{R} \ ,
\end{align}
we find
\begin{align}
8\pi G_\text{N} (T_{ww}+T_{tt} ) =\frac{d-2}{2w^2\sqrt{f(w)}} \left( \frac{w^2 f'(w)}{\sqrt{f(w)}}\right)'  \ge 0 \ .
\end{align}
Since $f(w)>0$ for $w>0$, we obtain the inequality
\begin{align}
f'(w) \geq 0 \ ,
\end{align}
which implies $f(w) \ge 1$ for $w>0$ or equivalently
\begin{align}
	\ell < 1 \ .
\end{align}

In this model, it follows from \eqref{Hol_EE_DCFT} and \eqref{area} that the difference of the defect entropies between the UV and IR fixed points is
\begin{align}
S_{\text{defect}}|_\text{UV} - S_{\text{defect}}|_\text{IR} =(1-\ell^{d-2})\,S_{\text{defect}}|_\text{UV} \ ,
\end{align}
which suggests the monotonicity of the  regularized defect entropy if the regularized value at the UV fixed point is positive.
On the other hand, one cannot calculate the defect free energy without specifying the bulk action that allows the domain wall metric as a solution.
Hence in what follows, we consider more explicit models and examine our proposal for the monotonicity of the defect free energy.

\subsection{Probe brane model}
As a concrete holographic model of DCFT we consider a brane system embedded in the AdS space.
In Euclidean signature, the action of the system becomes
\begin{align}
I_{d,p}=I_{\mathrm{EH}}+I_{\mathrm{brane}} \ ,
\label{73}
\end{align}
where $I_{\mathrm{EH}}$ is Einstein-Hilbert action with a cosmological constant
\begin{align}
I_{\mathrm{EH}}=-\frac{1}{16\pi G_{\text{N}}} \int_\CB \! \dd ^{d+1}X \, \sqrt{G}\left( \CR +\frac{d(d-1)}{L^2}\right)\ ,
\end{align}
and $I_{\mathrm{brane}}$ is a brane action
\begin{align}
I_{\mathrm{brane}} = T_p\,\int_\CQ \! \dd^{p+1}\xi \sqrt{\hat{G}}\ .
\end{align}
with the brane tension $T_p$ and the induced metric $\hat{G}_{AB}$ on the brane.
The bulk spacetime $\CB$ is fixed by solving the Einstein equation with the source from the brane on $\CQ$ which is anchored on the defect of dimension $p$ on the boundary $\CM \equiv \partial\CB$.

When the tension is small, $T_p L^{p+1}\ll 1$, the brane can be treated as a probe. 
In this limit, the defect free energy is given by minus the on-shell action of the brane
\begin{align}
	 \log\, \langle \, \CD^{(p)}\,\rangle = - I_\text{brane}\ .
\end{align}
The on-shell action is simply the volume of the brane times the brane tension
\begin{align}
	\begin{aligned}
	I_\text{brane} &= \text{Vol} (\BH^{p+1}) \, T_p L^{p+1}\\
    	&= - \frac{1}{\sin (\pi p/2)}\,\frac{\pi^{p/2 +1}}{\Gamma (p/2 + 1)}\, T_p L^{p+1} \ .
	\end{aligned}
\end{align}

We can similarly compute the leading contribution to the defect entropy in the probe limit.
For the spherical entangling region one finds \cite{Jensen:2013lxa}
\begin{align}\label{EE_probe}
	S_\text{defect}= \frac{1}{\sin \left( \pi p/2 \right)}\, \frac{p}{d-1+\delta_{pd}}\, \frac{\pi^{p/2 +1}}{\Gamma (p/2 +1)}\, T_p L^{p+1} \ .
\end{align}

It is worthwhile to pointing out that the defect entropy is proportional to the on-shell action
\begin{align}\label{S1_Ibrane}
	S_\text{defect} = - \frac{p}{d-1 + \delta_{pd}}\, I_\text{brane} \ .
\end{align}
Moreover they coincide up to the sign when $p=d-1$.
This should be compared with our field-theoretical result \eqref{Defect_Entropy_universal} relating the defect entropy to the on-shell action
\begin{align}\label{DE_Relative}
	S_\text{defect} = - I_\text{brane} - \frac{1}{\sin \left( \pi p/2 \right)}\, \frac{2(d-p-1)}{d\,\Gamma \left( (d-p)/2\right)}\, \frac{\pi^{d/2 +1}}{\Gamma (p/2 +1)}\,a_T \ ,
\end{align}
Comparing \eqref{S1_Ibrane} with \eqref{DE_Relative} we can read off $a_T$ for $p<d-1$ in the probe brane model
\begin{align}
	a_T = \frac{d}{2(d-1)\,\pi^{(d-p)/2}}\,\Gamma \left( \frac{d-p}{2}\right) \, T_p L^{p+1}\ .
\end{align}

In the case of a codimension-one defect ($p=d-1$), the backreacted metric takes the same form as \eqref{deform} with the range $-\infty < \rho < \infty$ and \cite{Chang:2013mca,Jensen:2013lxa}
\begin{align}
	A(\rho) = \cosh ( |\rho| - \rho_\ast ) \ , \qquad \rho_\ast \equiv \text{arctanh} \left(\frac{4\pi G_\text{N}\,T_{d-1}\,L}{d-1} \right) \ .
\end{align}
The defect entropy is given exactly by
\begin{align}\label{Hol_ICFT}
	S_\text{defect} 
    =  \frac{L^{d-1}}{2G_\text{N}} \,\frac{\pi^{(d-1)/2}}{\sin \left(\pi (d-1)/2\right)\,\Gamma\left( (d-1)/2\right) }\, \tanh \rho_\ast \cdot {}_2F_1 \left(\frac{1}{2}, \frac{d}{2}, \frac{3}{2}; \tanh^2 \rho_\ast \right) \ .
\end{align}
It reproduces \eqref{EE_probe} in the probe limit $\rho_\ast \to 0 \, (T_{d-1} L^{d}\ll 1)$ as expected.\footnote{This is twice the boundary entropy \eqref{Hol_Bdy_Entropy} calculated in the holographic model of BCFT in a later subsection.}

One can read off the universal part of the defect free energy in the probe brane model
\begin{align}\label{D_probe}
	\begin{aligned}
	\tilde D_\text{brane} &\equiv -\sin (\pi p/2)\, I_\text{brane} \\
	    &= \frac{\pi^{p/2 +1}}{\Gamma (p/2 +1)}\, T_p L^{p+1}\ ,
    \end{aligned}
\end{align}
which is seen to be positive for $T_p >0$.
Hence our conjecture \eqref{D_inequality} asserts that the brane tension must decrease under any defect RG flow.
This conforms to an intuition that the smaller the brane tension is, the less the degrees of freedom live on the defect (as there are no defects when $T_p=0$).
We will show the brane tension monotonically decreases under a defect RG flow described by a holographic model generalizing the probe brane model in the next subsection.

\subsection{A holographic model of defect RG flow}
We adopt a simple holographic model of defect CFT described by the same type of the action as \eqref{73} with $I_\text{brane}$ replaced by the action of a single real scalar field $\phi$ \cite{Yamaguchi:2002pa}
\begin{align}
	I_\text{brane} = \int \d^{p+1} \xi\, \sqrt{\hat{G}} \left[\frac{1}{2}\, \hat{G}^{AB}\partial_A \phi\, \partial_B \phi + V(\phi) \right] \ ,
\end{align}
on a $(p+1)$-dimensional hyperbolic space anchored on a $p$-dimensional defect at the boundary of the Euclidean AdS$_{d+1}$ space.
We assume that the potential $V(\phi)$ is bounded from below and allows a few critical points satisfying
\begin{align}
	\frac{\d V}{\d \phi} = 0 \ .
\end{align}
At each critical point $\phi_0$ this model reduces to the probe brane model with the brane tension
\begin{align}\label{Tension_Potential}
	T_p = V(\phi_0) \ ,
\end{align}
and the defect RG flow is triggered by letting $\phi$ roll off from a local maximum to a local minimum of $V(\phi)$.

Now we focus on a holographic dual of a planer defect on $\BR^d$.
In the Poincar\'e coordinates 
\begin{align}
	\d s^2 = \d r^2 + \mathrm{e}^{-2r/L}\,\delta_{\mu\nu}\,\d x^\mu \d x^\nu \ ,
\end{align}
the brane action is localized at $x^p = x^{p+1} = \cdots = x^{d-1} = 0$.
The worldvolume coordinates $\xi^A$ can be chosen as 
\begin{align}
	\xi^a = x^a\quad (a = 0, \cdots, p-1) \ , \qquad \xi^p = r \ .
\end{align}
Let us define a function
\begin{align}
	T(\phi) \equiv V(\phi) - \frac{1}{2}\, (\partial_r \phi)^2 \ ,
\end{align}
then it is easy to show $T(\phi)$ is a monotonically decreasing function with respect to $r$ \cite{Yamaguchi:2002pa},
\begin{align}\label{T_monotone}
	\partial_r T(\phi) = -\frac{p}{L}\, (\partial_r \phi)^2 \le 0 \ ,
\end{align}
where we use the equation of motion of $\phi$ and the translation invariance of the solution along the defect. 
For the holographic RG flow interpolating between the UV fixed point $\phi_\text{UV}$ and the IR $\phi_\text{IR}$, \eqref{T_monotone} implies that the critical value of the potential is non-increasing under the RG flow,
\begin{align}
	V(\phi_\text{UV}) \ge V(\phi_\text{IR}) \ ,
\end{align}
which in turn yields the brane tension is non-increasing in the probe brane model
\begin{align}
	T_{p,\text{UV}} \ge T_{p,\text{IR}} \ .
\end{align}
With \eqref{D_probe} in mind we find the monotonicity
\begin{align}
	\tilde D_\text{brane}|_\text{UV} \ge \tilde D_\text{brane}|_\text{IR} \ ,
\end{align}
in accordance with our proposal \eqref{D_inequality}.

\subsection{AdS/BCFT model}
Finally we examine the $g$-theorem stating the monotonicity of the hemisphere partition function of BCFTs under any boundary RG flow.
The bulk AdS metric respecting the $\SO (1,d)$ symmetry of BCFT$_d$ on a hemisphere is
\begin{align}\label{BCFT_coord}
	\d s^2 = L^2 \left[\d \rho^2 + \cosh^2\rho \left( \d w^2 + \sinh^2 w\,\d s_{\mathbb{S}^{d-1}}^2 \right)\right] \ ,
\end{align}
where $\rho \in (-\infty, \infty)$ and $w\in (0, \infty)$.
This metric is equivalent to the more familiar form of the global AdS space
\begin{align}\label{BCFT_global}
	\d s^2 = L^2 \left[ \d u^2 + \sinh^2 u \left( \d\theta^2 + \cos^2\theta\, \d s_{\mathbb{S}^{d-1}}^2 \right)\right] \ ,
\end{align}
where $u\in (0, \infty)$ and $\theta \in [-\pi /2, \pi/2]$.
They are related by the following coordinate transformation
\begin{align}\label{BCFT_coord_tr}
	\cot \theta = \coth \rho\, \sinh w \ , \qquad \cosh u = \cosh \rho\, \cosh w \ .
\end{align}
The hemisphere defined by $\theta \in [-\pi/2, 0]$ at $u=\infty$ is reached by the $\rho \to -\infty$ limit in the coordinates \eqref{BCFT_coord} while the other half defined by $\theta \in [0, \pi/2]$ at $u=\infty$ is reached by the $\rho \to \infty$ limit.
The boundary of the hemisphere at $\theta = 0$ is reached by the $w\to \infty$ limit for any $\rho$.

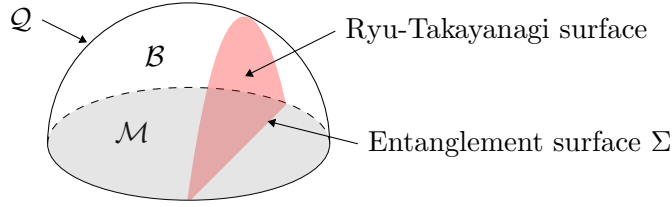
\begin{figure}[t]
  \centering
\begin{tikzpicture}[scale=1.5]
   \draw[dashed] (4.25,1) arc [start angle = 0, end angle = 180, x radius=1.25cm, y radius=0.5cm];
   \draw (4.25,1) arc [start angle = 0, end angle = -180, x radius=1.25cm, y radius=0.5cm];
   \draw (4.25,1) arc [start angle = 0, end angle = 180, radius = 1.25cm];
   \draw (2.5,1.1) node {$\mathcal{M}$};
   \draw (2.7,1.6) node [above] {$\mathcal{B}$};
   \draw[-{Triangle[angle'=60,scale=0.8]}] (1.7,2.1) node [left] {$\mathcal{Q}$} --(2.1,1.9);
   \filldraw[  fill=red,opacity=0.3,draw=red,line width=1pt] (3,0.5)--(3.85,1.35).. controls (3.85,1.35) and (3.425,3.55) .. (3,0.5);
   \filldraw[   gray,opacity=0.2,draw=none] (3,1) circle [x radius = 1.25cm, y radius=0.5cm];
   \draw[{Triangle[angle'=60,scale=0.8]}-] (3.5,1.6)--(4.3,2.0) node [right] {Ryu-Takayanagi surface};
   \draw[{Triangle[angle'=60,scale=0.8]}-] (3.7,1.2)--(4.5,1.0) node [right] {Entanglement surface $\Sigma$};
\end{tikzpicture}
   \caption{The bulk AdS space $\mathcal{B}$ is surrounded by the union $\mathcal{M} \cup \mathcal{Q}$ of the boundary and bulk hemispheres. It is bipartited by the Ryu-Takayanagi surface anchored on the entanglement surface $\Sigma$.}
   \label{fig:AdSBCFT}
\end{figure}

We locate BCFT on the hemisphere covered by $\theta \in [-\pi/2, 0]$ and construct the gravity dual following Takayanagi's proposal \cite{Takayanagi:2011zk,Fujita:2011fp,Nozaki:2012qd} by introducing the AdS boundary $\CQ$ with a brane of tension $T$,
\begin{align}
	\begin{aligned}
	I &= - \frac{1}{16\pi G_\text{N}} \int_{\CB}\sqrt{G}\, \left( \CR + \frac{d(d-1)}{L^2}\right) \\
    	&\qquad - \frac{1}{8\pi G_\text{N}} \int_{\CQ}\sqrt{\hat{G}} \left( \CK - T\right) - \frac{1}{8\pi G_\text{N}} \int_{\CM}\sqrt{\hat{G}}\, \CK \ ,
\end{aligned}
\end{align}
where $\CB$ is the bulk AdS space and $\CM$ is the boundary on which the dual BCFT lives.
In the present case, $\CM$ is the hemisphere, $\CB$ is the bulk AdS space in the coordinates \eqref{BCFT_coord} with the restricted range $\rho \in (-\infty , \rho_\ast)$, and $\CQ$ is the AdS boundary at $\rho = \rho_\ast$ (see figure \ref{fig:AdSBCFT}).
To make the variational problem well-defined in the presence of the boundary, the Gibbons-Hawking term is introduced with the extrinsic curvature defined by 
\begin{align}
	\CK_{MN} = \hat{G}_{ML}\hat{G}_{NK}\nabla^L n^K \ ,
\end{align}
for the outward pointing normal vector $n^M$.
The Dirichlet boundary condition is imposed on $\CM$, but the Neumann boundary condition is chosen on $\CQ$
\begin{align}
	\CK_{MN} -\hat{G}_{MN}\,\CK = - T\, \hat{G}_{MN} \ .
\end{align}
Since the extrinsic curvature is given by
\begin{align}
	\CK = \frac{d}{L}\,\tanh \rho \ ,
\end{align}
for any constant $\rho$ surface, the brane tension is fixed to be
\begin{align}
	T = \frac{d-1}{L}\,\tanh \rho_\ast \ .
\end{align}

\subsubsection{On-shell action at critical points}
We now calculate the on-shell action of this system.
Without regularization, the on-shell action diverges,
\begin{align}\label{Diff_on_shell}
		I (\rho_\ast) = \frac{L^{d-1}}{8\pi G_\text{N}}\, \text{Vol}(\BH^d)\,\left[ d\, \int_{-\infty}^{\rho_\ast}\d \rho\,\cosh^d \rho - \frac{T L}{d-1}\,\cosh^d \rho_\ast + d \lim_{\rho\to -\infty} \tanh \rho\, \cosh^d \rho\right] \ .
\end{align}
To compare with the boundary entropy \eqref{Boundary_Entropy_universal} we subtract half of the on-shell action $I_\text{AdS}$ of the whole AdS space without branes, which equals the on-shell action \eqref{Diff_on_shell} with $\rho =0$ where the brane tension vanishes,
\begin{align}
	\frac{1}{2}\,I_\text{AdS} = I(0) \ .
\end{align}
Hence the boundary entropy \eqref{Boundary_Entropy_universal} reads
\begin{align}\label{BE_onshell}
	\begin{aligned}
		S_\text{bdy} &= -I (\rho_\ast) + \frac{1}{2}\,I_\text{AdS} \\
        &= \frac{L^{d-1}}{4 G_\text{N}}\, \frac{\pi^{(d-1)/2}}{\sin\left( \pi (d-1)/2\right)\,\Gamma \left( (d-1)/2\right)}\,\tanh \rho_\ast \cdot {}_2F_1 \left(\frac{1}{2}, \frac{d}{2}, \frac{3}{2}; \tanh^2\rho_\ast \right) \ .
	\end{aligned}
\end{align}
We can read off the universal part of the boundary entropy from \eqref{D_coefficient_bdy},
\begin{align}\label{D_BCFT}
	\tilde D (\rho_\ast) =  \frac{L^{d-1}}{4 G_\text{N}}\, \frac{\pi^{(d-1)/2}}{\Gamma \left( (d-1)/2\right)}\,\tanh \rho_\ast \cdot {}_2F_1 \left(\frac{1}{2}, \frac{d}{2}, \frac{3}{2}; \tanh^2\rho_\ast \right) \ ,
\end{align}
which can be checked numerically to be a monotonically increasing function of $\rho_\ast$.

\subsubsection{Holographic boundary entropy}
We now calculate the entanglement entropy of a half ball region in BCFTs following \cite{Jensen:2013lxa}.
Let $r_\perp$ be the transverse coordinate to the boundary and 
introduce the metric in the flat space
\begin{align}
	\d s^2 = \d t^2 + \d r_{||}^2 + r_{||}^2\, \d s_{\mathbb{S}^{d-3}}^2 + \d r_\perp^2 \ .
\end{align}
BCFTs are defined in the domain $r_\perp \in [0, \infty)$ and the entangling surface is located at the hypersurface satisfying $r_\perp^2 + r_{||}^2 = R^2$ at $t=0$.

The gravity dual of the BCFT is described by the hyperbolic slicing of the AdS$_{d+1}$ spacetime
\begin{align}
	\d s^2 = L^2 \left[ \d \rho^2 + \cosh^2 \rho\, \frac{\d z^2 + \d t^2 + \d r_{||}^2 + r_{||}^2\, \d s_{\mathbb{S}^{d-3}}^2}{z^2}\right] \ ,
\end{align}
where $\rho \in (-\infty, \rho_\ast)$ and $z\in [0, \infty)$.
The holographic entanglement entropy is given by the area of the Ryu-Takayanagi surface parametrized by $z(x, r_{||})$ at $t=0$, but assuming the $x$-independence one finds the unique semi-circle solution \cite{Jensen:2013lxa} (see figure \ref{fig:AdSBCFT})
\begin{align}
	z(r_{||})^2 + r_{||}^2 = R^2 \ .
\end{align}
Then we find the holographic boundary entropy 
\begin{align}\label{Hol_Bdy_Entropy}
	\begin{aligned}
		S_\text{bdy} &= \frac{L^{d-1}}{4G_\text{N}}\,\text{Vol}(\BH^{d-2})\, \left( \int_0^{\rho_\ast} \d \rho\, \cosh^{d-2}\rho\right) \\
        	&= \frac{L^{d-1}}{4G_\text{N}}\, \frac{\pi^{(d-1)/2}}{\sin\left( \pi(d-1)/2\right)\, \Gamma\left( (d-1)/2\right)}\, \tanh\rho_\ast\cdot {}_2F_1 \left(\frac{1}{2}, \frac{d}{2}, \frac{3}{2}; \tanh^2\rho_\ast \right) \ .
	\end{aligned}
\end{align}
This is equivalent to the boundary entropy \eqref{BE_onshell} calculated by the formula \eqref{Boundary_Entropy_universal} with the on-shell action as expected.

\subsubsection{Holographic $g$-theorem}
We shall prove the holographic $g$-theorem in the AdS/BCFT model by adapting the setup of \cite{Takayanagi:2011zk,Fujita:2011fp} to the present case.
Since the on-shell action \eqref{Diff_on_shell} is a monotonic function of the brane tension the proof amounts to showing the monotonicity of the brane tension under boundary RG flows.
The strategy of the proof is in parallel with the holographic $C$-theorem \cite{Freedman:2003ax} where the null energy condition is imposed on the bulk matter field to construct a monotonic function of a bulk metric component with respect to the holographic coordinate.
In the present case, we would rather impose the null energy condition on the boundary $\CQ$ for any null vector $\zeta^M$
\begin{align}\label{NEC_BCFT}
	(\CK_{MN} - \CK\, \hat{G}_{MN})\, \zeta^M \zeta^N \ge 0 \ .
\end{align}
For a boundary RG flow respecting the $\SO(d)$ symmetry on the boundary of the hemisphere, the brane configuration on $\CQ$ is fixed by 
\begin{align}
	\theta = \theta (u) \ ,
\end{align}
in the global AdS coordinates \eqref{BCFT_global}.
To impose the null energy condition, 
we analytically continue \eqref{BCFT_global} to the Lorentzian signature by replacing the boundary sphere with the de Sitter space,
\begin{align}
	\d s_{\mathbb{S}^{d-1}}^2~ \longrightarrow ~ - \d t^2 + \cosh^2 t\, \d s_{\mathbb{S}^{d-2}}^2 \ .
\end{align}
The resulting metric becomes
\begin{align}
	\d s^2 = L^2 \left[ \d u^2 + \sinh^2 u \left( \d\theta^2 -\cos^2\theta\,\d t^2 + \cos^2\theta\,\cosh^2 t\, \d s_{\mathbb{S}^{d-2}}^2\right)\right] \ ,
\end{align}
in which the outward pointing unit normal vector to $\CQ$ is given by
\begin{align}
	n^u = - \frac{L\,\theta'(u)}{\sqrt{ \theta'(u)^2 + \text{csch}^2 u}}\ , \qquad n^\theta = \frac{L}{\sqrt{ \theta'(u)^2 + \text{csch}^2 u}}  \ ,\qquad n^{M\neq u,\theta} = 0 \ .
\end{align}
Choosing the null vector to be
\begin{align}
	\zeta^u = \text{const} \ , \qquad \zeta^\theta = \zeta^u\,\theta'(u) \ ,\qquad \zeta^t = \zeta^u\,\frac{\sqrt{ \theta'(u)^2 + \text{csch}^2 u}}{\cos \theta}\ , \qquad \zeta^{M\neq u, \theta, t} = 0 \ ,
\end{align}
the condition \eqref{NEC_BCFT} becomes
\begin{align}
	\frac{L\, (\zeta^u)^2}{\sqrt{ \theta'(u)^2 + \text{csch}^2 u}}\, g(u) \ge 0 \ ,
\end{align}
where
\begin{align}
	g(u) \equiv  \tan \theta(u)\,\text{csch}^2 u - \theta'(u)\,\coth u  + \theta'(u)^2\, \tan\theta(u) - \theta''(u) \ .
\end{align}
Hence the null energy condition yields the non-negativity of the function 
\begin{align}
	g(u)\ge 0 \ .
\end{align}

Next we want to show the monotonicity of the brane angle $\rho_\ast$.
We switch to the coordinates \eqref{BCFT_coord} where the brane is located on the hypersurface
\begin{align}
	\rho (u) = \text{arcsinh} \left( \sinh u\, \sin \theta (u) \right) \ .
\end{align}
In what follows we show the derivative is non-negative
\begin{align}\label{rho_monotonic}
	\rho'(u) = \frac{1}{\sqrt{ \theta'^2 + \text{csch}^2 u}} \, f(u) \ge 0 \ ,
\end{align}
where 
\begin{align}
	f(u) \equiv \sin \theta(u)\, \coth u + \theta'(u) \cos\theta(u) \ .
\end{align}
As long as the brane configuration satisfies $0\le \theta(u) \le \pi/2$ the null energy condition implies
\begin{align}
	f'(u) = - \cos \theta(u)\, g(u) \le 0 \ .
\end{align}
As the boundary condition $\lim_{u\to \infty} \theta(u) = 0$ imposes $f(\infty) = 0$ we conclude $f(u)\ge 0$ and $\rho'(u) \ge 0$ for $u \in [0, \infty)$.

The inequality \eqref{rho_monotonic} means the brane angle $\rho_\ast$ monotonically decreases under a boundary RG flow
\begin{align}
	\rho_\text{UV} \ge \rho_\text{IR} \ ,
\end{align}
where $\rho_\text{UV} = \rho (\infty)$, $\rho_\text{IR} = \rho(0)$, and we interpret the coordinate $u$ as the holographic renormalization scale as in \cite{Freedman:2003ax}.
Combined with \eqref{D_BCFT} at the critical point, we prove the weak form of the holographic $g$-theorem
\begin{align}
	\tilde D_\text{UV} \ge \tilde D_\text{IR} \ .
\end{align}

\section{Discussion}
\label{con}

The $C$-theorems in BCFTs and DCFTs were proposed  in various forms and
one of the purposes of this paper was to organize the previous studies scattered in literatures.
Along the way we found some of the conjectures turned out to be equivalent.
In pursuing the unified picture of the $C$-theorems,
we proposed that the defect free energy be a $C$-function in DCFTs of any dimensions with any dimensional defects.
We tested our proposal both in field theory and holographic models.
In field theory side, we considered Wilson loops as line defects and provided a few supporting evidences.
As a by-product, we revealed that the defect entropy does not always decrease under a defect RG flow.
Furthermore, we were able to prove our conjecture in various holographic models describing DCFTs.

While concrete examples of the defect (boundary) RG flow in DCFTs (BCFTs) are less known so far, our conjecture has been proven as the $b$-theorem when $p=2$ non-perturbatively \cite{JO1}.
Thus it is intriguing to examine if the proof of the $b$-theorem can be extended to the higher-dimensional cases.
Also there may be an analogue of the $F$-maximization \cite{Closset:2012vg} in supersymmetric DCFTs as suggested by \cite{Gaiotto:2014gha}.
We leave these problems for future investigations. 

The $C$-theorem \eqref{D_inequality}  we propose is in a weak form, and it is reasonable to ask whether there exists a strong (or even stronger) version of the $C$-theorem that requires the monotonicity of the $C$-function along the entire defect RG flow (and the stationarity at the fixed point).
For instance, the $g$-function in BCFT$_2$ built from the relative entropy is a strong $C$-function \cite{Casini:2016fgb}.
Hence one may hope to extend such a construction to higher dimensions and establish the strong version of our proposal.
In appendix \ref{ap:Relative} 
we considered the relative entropy between DCFT and CFT for a spherical entangling region and showed the equivalence to the defect free energy via the CHM map.
We further derived the inequality \eqref{Relative_monotone} from the positivity of the relative entropies, which suggests the monotonicity theorem \eqref{DF_diff_RE} for the \emph{unrenormalized} defect free entropy.
Since the defect free energy and the relative entropy have UV divergences of the order of $O(\epsilon^{-p})$ for a $p$-dimensional defect, \eqref{DF_diff_RE} does not prove our conjecture but rather derives the monotonicity of the leading coefficient proportional to the volume, which reminds us of the area theorem for the entanglement entropy \cite{Casini:2016udt, Casini:2017vbe}.
It is tempting to improve this method, possibly with the monotonicity of the relative entropy, for proving the strong form of our conjecture.

In the examples  we considered in section \ref{sec3}, we showed that the monotonicity of the defect entropy failed when the rank $N$ of the gauge group is small. 
On the other hand, the holographic models correspond to the large $N$ limit and both the defect free energy and the defect entropy decrease under the defect RG flows.
If the $1/N$ correction is taken into account in the holographic entanglement entropy following the Faulkner-Lewkoywcz-Maldacena procedure \cite{Faulkner:2013ana}, one may be able to observe the violation of the monotonicity of the defect entropy even in the holographic models we considered.

\acknowledgments
We would like to thank T.\,Okuda, S.\,Sasa, T.\,Ugajin and Y.\,Zhou for useful discussion.
The work of N.\,K. was supported in part by the Program for Leading Graduate Schools, MEXT, Japan and also supported by World Premier International Research Center Initiative (WPI Initiative), MEXT, Japan.
This work of T.\,N. was supported in part by the JSPS Grant-in-Aid for Young Scientists (B) No.15K17628 and the JSPS Grant-in-Aid for Scientific Research (A) No.16H02182.
The work of Y.\,S. is supported by the Grant-in-Aid for Japan Society for the Promotion of Science Fellows, No.16J01567.
The work of K.\,W. is supported by the Grant-in-Aid for Japan Society for the Promotion of Science Fellows, No.18J00322.

\appendix

\section{Terminology and notation} 
\label{notation}

We summarize our terminology and notation used in this paper.

\begin{itemize}
\item An ambient space where the CFT lives is $d$-dimensional and is labeled by the Greek letters $\mu,\nu,\cdots$. It is common to use ``ambient" instead of ``bulk" in BCFT or DCFT literatures to avoid a confusion.
\item A defect is $p$-dimensional, whose worldvolume coordinates are labeled by the Roman letters $a,b,\cdots$.
The quantities on the defect are hatted to distinguish from the ambient ones.
For example, a scalar operator localized on the defect is denoted by $\hat{\mathcal{O}}(\hat{x})$. 
\item The transverse directions to the defect are labeled by the Roman indices $i,j,\cdots$.
\item A bulk space holographically dual to DCFT is $(d+1)$-dimensional whose coordinates are labeled by the capital Roman letters $M,N,\cdots$.
\item
In some holographic models, the defect is introduced by a brane in the bulk. 
The coordinates on the branes are labeled by the capital Roman letters $A,B,\cdots$.
\end{itemize}

\paragraph*{Coordinate}

In the field theory side, we split the coordinates $x^\mu=(\hat{x}^a,x_\perp^i)$ where
\begin{itemize}
\item $x^\mu$ : the ambient space coordinate,
\item $\hat{x}^a$ : the defect worldvolume coordinate,
\item $x_\perp^i$ : the coordinates transverse to the defect.
\end{itemize}

\noindent
In the holography side, we use
\begin{itemize}
\item $X^M$ : the bulk (AdS) space coordinates,
\item $\xi^A$ : the brane coordinates.
\end{itemize}

\paragraph*{Metric}
To distinguish the metrics in the ambient, defect worldvolume, bulk and brane worldvolume coordinates, we use
\begin{itemize}
\item $g_{\mu \nu}$ : the ambient space metric,
\item $\hat{g}_{ab}=\frac{\partial x^\mu}{\partial x^a}\frac{\partial \hat{x}^\nu}{\partial \hat{x}^b}g_{\mu \nu}$ : the defect worldvolume metric (the induced metric on the defect),
\item $G_{MN}$ : the bulk (AdS) space metric,
\item $\hat{G}_{AB}$ : the induced metric on the brane in holographic models 
\end{itemize}

\section{Relative entropy in DCFT and defect free energy}\label{ap:Relative}
The relative entropy between two states $\rho$ and $\sigma$ is a measure of distinguishability defined by
\begin{align}
	S (\,\rho\, ||\, \sigma\,) \equiv \tr \left[ \rho\, \log \rho\right] -\tr \left[ \rho\,\log \sigma\right] \geq 0 \ .
\end{align}
It is non-negative and vanishes if and only if $\rho=\sigma$.
It is monotonic under the inclusion of the subsystems $A \supseteq \tilde{A}$ or their causal domains $\mathcal{D}(A) \supseteq \mathcal{D}(\tilde{A})$, 
\begin{align}
S (\,\rho_{A}\, ||\, \sigma_{A}\,) \geq S (\,\rho_{\tilde{A}}\, ||\, \sigma_{\tilde{A}}\,) \ .
\end{align}
It is also stationary around the reference state $\sigma$, that is,   
for perturbed states $\rho = \sigma + \delta \rho$, 
\begin{align}
S (\sigma + \delta \rho\, ||\, \sigma ) = -\frac{1}{2} \,\mathrm{tr} \left[ \delta \rho \sigma^{-1} \delta \rho \right] + O(\delta \rho^3) \ .
\end{align}
where $\mathrm{tr} \left[\delta \rho \right] =0$. 

The additional entanglement due to defects can be measured by the relative entropy between DCFT and CFT,
\begin{align}\label{Defect_Relative}
	S (\,\rho^\text{(DCFT)}\, ||\, \rho^\text{(CFT)}\,) = \Delta \langle \, H_\text{CFT}\, \rangle - S_\text{defect} \ ,
\end{align}
where $H = -\log \rho$ is the modular Hamiltonian 
for the (normalized) reduced density matrix ($\text{tr}\, \rho =1$) and
\begin{align}
	\Delta \langle \, H_\text{CFT}\, \rangle \equiv  \langle \, H_\text{CFT}\, \rangle^\text{(DCFT)} - \langle \, H_\text{CFT}\, \rangle^\text{(CFT)} \ .
\end{align}
The modular Hamiltonian is non-local in general, but takes a simple form for a spherical entangling surface in CFT as it generates the translation along $\tau$ direction in the hyperbolic coordinates \eqref{Hyperbolic},
\begin{align}
	H_\text{CFT} = \int_{\BS^{1} \times \BH^{d-1}}\, (T_\text{CFT})^{\tau}_{~\tau} + S^\text{(CFT)}\ .
\end{align}
The constant part $S^\text{(CFT)}$ is fixed by taking the vev of both sides in CFT with $\langle\, (T_\text{CFT})_{\mu\nu}\,\rangle_\text{CFT} = 0$.
The constant term of the modular Hamiltonian in DCFT can also be fixed by \eqref{I_vev} and \eqref{Defect_Entropy_bare},
\begin{align}
	H_\text{DCFT} = \int_{\BS^{1} \times \BH^{d-1}}\, (T_\text{DCFT})^{\tau}_{~\tau} + \log Z^\text{(DCFT)}\ .
\end{align}
Here one must use the relation $S^\text{(CFT)} = \log Z^\text{(CFT)}$ that holds up to UV divergences.

To simplify the discussion, we restrict our attention to the odd-dimensional case, but it should be straightforward to generalize the following argument to the even-dimensional case.
In CFT without defects, the one-point function of the stress tensor vanishes (up to a constant term) while it does not in defect CFTs.
Thus we find 
\begin{align}\label{Modular_Hamiltonian_diff}
	\Delta \langle \, H_\text{CFT}\, \rangle = \int_{\BS^{1} \times \BH^{d-1}}\, \langle \,(T_\text{CFT})^{\tau}_{~\tau}\,\rangle^\text{(DCFT)}\ .
\end{align}
Combined with \eqref{Defect_Entropy_bare} and \eqref{Defect_Relative}, we find
\begin{align}
	S (\,\rho^\text{(DCFT)}\, ||\, \rho^\text{(CFT)}\,) = - \log\, \langle\, \CD^{(p)}\,\rangle \ ,
\end{align}
where we use the fact \eqref{Defect_ST_vanish} that the vev of the defect localized stress tensor vanishes $\langle \,t^{\tau}_{~\,\tau}\, \rangle^\text{(DCFT)}=0$.
Note that this is the relation between the universal parts in the both side, so
the positivity of the relative entropy does not necessarily mean the (unrenormalized) defect free energy is negative.

The relative entropy contains a stronger UV divergent term than the defect entropy
\begin{align} \label{RE_UVdiv}
	S (\,\rho^\text{(DCFT)}\, ||\, \rho^\text{(CFT)}\,) = \frac{c_{p}}{\epsilon^{p}} + \frac{c_{p-2}}{\epsilon^{p-2}} + \cdots +
    	\begin{cases}
    		B \log \epsilon + \cdots \ , & (p: \text{even}) \ , \\
            (-1)^{(p+1)/2}\,D \ ,  & (p: \text{odd}) \ .
    	\end{cases}
\end{align}
The difference between the two entropies follows from the relation \eqref{Defect_Relative}, where the modular Hamiltonian term \eqref{Modular_Hamiltonian_diff} gives the UV divergences of order $O(\epsilon^{-p})$.
To see this, we excise the tubular neighborhood of radius $\epsilon$ of the defect and integrate the stress tensor \eqref{Stress_tensor_One_Point} from $x=\epsilon$ to $\infty$,
\begin{align}
	\begin{aligned}
	\Delta \langle\, H_\text{CFT}\,\rangle &\sim \int_\epsilon^\infty \d x\, \frac{a_T}{\sinh^{d} x}\, \cosh^{p-1}x\,\sinh^{d-p-1}x 
    \sim \frac{a_T}{\epsilon^p} \ .
    \end{aligned}
\end{align}
This explains the leading UV divergence in \eqref{RE_UVdiv}.

Next let us turn to the relevant perturbation of DCFT parametrized by a relevant coupling $\hat{\lambda}$. 
We consider the relative entropy between a perturbed state in DCFT whose reduced density matrix is denoted by $\rho^\text{(DCFT)}_{\hat{\lambda}}$ and a vacuum state in CFT.
Then the difference of the relative entropies between the perturbed and unperturbed states becomes
\begin{align}
	\begin{aligned}
S (\,\rho^\text{(DCFT)}_{\hat{\lambda}}\, &||\, \rho^\text{(CFT)}\,)  - S (\,\rho^\text{(DCFT)}\, ||\, \rho^\text{(CFT)}\,) \\
	& = - S_{\hat{\lambda}}^\text{(DCFT)} + \langle\, H_\text{CFT} \,\rangle^\text{(DCFT)}_{\hat{\lambda}} + S^\text{(DCFT)} - \langle\, H_\text{CFT}\, \rangle^\text{(DCFT)} \\
 & =  S  (\,\rho^\text{(DCFT)}_{\hat{\lambda}}\, ||\, \rho^\text{(DCFT)}\,) - \langle\, h\,\rangle^\text{(DCFT)}_{\hat{\lambda}} + \langle\, h\, \rangle^\text{(DCFT)} \ ,
	\end{aligned}
\end{align}
where we introduced 
\begin{align}
	h \equiv \int_{\BS^{1} \times \BH^{d-1}}\,t^{\tau}_{~\tau} \ .
\end{align}
This term may be ignored if one employs the regularization scheme such that one excises the small tubular neighborhood of the defect.
In this case, using the positivity of $S  (\,\rho^\text{(DCFT)}_{\hat{\lambda}}\, ||\, \rho^\text{(DCFT)}\,)$, we can observe that the strong form of the defect RG flow monotonicity for $S (\,\rho^\text{(DCFT)}_{\hat{\lambda}}\, ||\, \rho^\text{(CFT)}\,)$ holds 
\begin{align}\label{Relative_monotone}
S (\,\rho^\text{(DCFT)}_{\hat{\lambda}}\, ||\, \rho^\text{(CFT)}\,) & \geq S (\,\rho^\text{(DCFT)}\, ||\, \rho^\text{(CFT)}\,) \ .
\end{align}
Alternatively, we can consider another rearrangement of the relative entropy
\begin{align}
	\begin{aligned} \label{DCFT_diff_RE}
		 S  (\,\rho^\text{(DCFT)}_{\hat{\lambda}}\, ||\, \rho^\text{(DCFT)}\,) 
            &= S (\,\rho^\text{(DCFT)}_{\hat{\lambda}}\, ||\, \rho^\text{(CFT)}\,)  + \langle\, h\,\rangle^\text{(DCFT)}_{\hat{\lambda}} + \log\, \langle\, \CD^{(p)}\,\rangle \ .
	\end{aligned}
\end{align}
When the perturbed DCFT flows to the IR fixed point, the first two terms in the right hand side becomes (see \eqref{Defect_Entropy_bare} and \eqref{Defect_Relative})
\begin{align}
\begin{aligned}
\left. S (\,\rho^\text{(DCFT)}_{\hat{\lambda}}\, ||\, \rho^\text{(CFT)}\,)  + \langle\, h\,\rangle^\text{(DCFT)}_{\hat{\lambda}} \right|_{\hat{\lambda} \to \hat{\lambda}_{\mathrm{IR}}} = \left. -\log\, \langle\, \CD^{(p)}\,\rangle \right|_{\mathrm{IR}}\ . 
\end{aligned}
\end{align}
Then the positivity of $S  (\,\rho^\text{(DCFT)}_{\hat{\lambda}}\, ||\, \rho^\text{(DCFT)}\,)$ in \eqref{DCFT_diff_RE} implies the weak form of the monotonicity for the unrenormalized defect free energy \eqref{Defect_Free_Energy} under the defect RG flow
\begin{align}\label{DF_diff_RE}
		\left. \log\, \langle\, \CD^{(p)}\,\rangle \right|_{\mathrm{UV}}
        \geq \left. \log\, \langle\, \CD^{(p)}\,\rangle \right|_{\mathrm{IR}}\ .
\end{align}
Note that this inequality only implies the monotonicity of the leading coefficient $c_p$ of the UV divergent terms in \eqref{UV_defect_FE}.
This resembles to the area theorem of the entanglement entropy \cite{Casini:2016udt, Casini:2017vbe}, but ours is much weaker statement than theirs.
It would be intriguing to derive a tighter inequality than \eqref{DF_diff_RE} by removing the UV divergences by extending the method of \cite{Casini:2016fgb} to the present case.

\bibliographystyle{JHEP}
\bibliography{Defect_Entropy}

\end{document}